\documentclass[aps,pre,twocolumn,superscriptaddress,amsmath,amssymb,longbibliography]{revtex4-1}

\def\bea{\begin{eqnarray}}
\def\eea{\end{eqnarray}}
\usepackage{graphicx}
\usepackage{dcolumn}
\usepackage{enumitem,kantlipsum}

\usepackage{bm}
\usepackage[normalem]{ulem}
\usepackage{color}
\usepackage{gensymb}
\usepackage{mathtools}
\usepackage{amsmath}

\usepackage{epstopdf}

\newcommand{\bew}{\begin{widetext}}
\newcommand{\ew}{\end{widetext}}
\newcommand{\nn}{\nonumber}

\newcommand{\bp}{\mathbf{p}}
\newcommand{\bq}{\mathbf{q}}

\newcommand{\bv}{\mathbf{v}}

\newcommand{\br}{\mathbf{r}}

\newcommand{\bn}{\mathbf{n}}

\newcommand{\sep}{ \ \ \ , \ \ \ }

\newcommand{\beq}{\begin{equation}}
\newcommand{\eeq}{\end{equation}}
\newcommand{\beqn}{\begin{eqnarray}}
\newcommand{\eeqn}{\end{eqnarray}}
\newcommand{\pp}{\partial}

\usepackage{xcolor}
\usepackage{color}

\definecolor{green}{rgb}{0,0.5,0}

\newcommand{\rv}{{\bf r}}

\newcommand{\E}{{\bf E}}

\newcommand{\bS}{{\bf S}}
\newcommand{\xv}{{\bf r}}

\newcommand{\qv}{{\bf q}}

\newcommand{\zh}{{\hat{\bf z}}}

\newcommand{\pv}{{\bf p}}

\newcommand{\kbt}{{k_BT}}

\newcommand{\oh}{{\frac{1}{2}}}

\newcommand{\grad}{{\bm{\nabla}}}

\newcommand{\tK}{\tilde K}
\newcommand{\bse}{\begin{subequations}}
\newcommand{\ese}{\end{subequations}}
\def\rf#1{(\ref{#1})}
\def\rfs#1{Eq.~\rf{#1}}

\begin{document}
\title{Coulomb universality}
\author{Leo Radzihovsky}
\affiliation{
Department of Physics and Center for Theory of Quantum Matter,
University of Colorado, Boulder, CO 80309}
\email{radzihov@colorado.edu}

\author{John Toner}
\affiliation{
  Department of Physics and Institute for Fundamental Science,
  University of Oregon, Eugene, OR 97403}
\email{jjt@uoregon.edu}
\date{\today}

\begin{abstract}
  Motivated by a number of realizations of long-range interacting
  systems, including ultra-cold atomic and molecular gases, we study a
  neutral plasma with power-law interactions {\it longer-ranged} than
  Coulombic.  We find that beyond a crossover length, such
  interactions are universally screened down to a standard Coulomb
  form in all spatial dimensions.  This implies, counter-intuitively,
  that in two dimensions and below, such a "super-Coulombic'' gas is
  asymptotically Coulombically confining at low temperatures. At
  higher temperatures, the plasma undergoes a deconfining transition
  that in two dimensions is the same Kosterlitz-Thouless transition
  that occurs in a conventional Coulomb gas, but at an elevated
  temperature that we calculate.  We also predict that in contrast,
  above two dimensions, even when naively the bare potential is
  confining, there is no confined phase of the plasma at any nonzero
  temperature. In addition, the super-Coulomb to Coulomb crossover is
  followed at longer length scales by an {\em unconventional}
  "Debye-Huckel" screening, which leads to faster-than-Coulombic,
  power-law decay of the screened potential, in contrast to the usual
  exponentially decaying Yukawa potential.  Furthermore, we show that
  power-law potentials that fall off more rapidly than Coulomb are
  screened down to a shorter-ranged power-law, rather than an
  exponential Debye-Huckel Yukawa form. We expect these prediction to
  be testable in simulations, and hope they will inspire experimental
  studies in various platforms.
\end{abstract}
 
\maketitle


\section{Introduction}
\subsection{Motivation} Recently there have been a number of
experimental realizations of long-range interacting systems,
particularly in ultra-cold atomic and molecular gases. These include
pseudo-spin systems of dipolar-interacting molecular
gases\cite{JunYe2013, Bakr2022}, trapped ions\cite{SageRMP2019} and
Rydberg atoms\cite{Browaeys2022}.  Low-dimensional condensed matter
surface systems coupled to a gapless, higher dimensional "bulk'' also
display long-range generalized elasticity\cite{surfaceLRnematicPRL,
  surfaceLRnematicPRE, surfaceLRsmecticEuroLett, surfaceLRsmecticPRE},
which lead to power-law interacting topological defects. 

Motivated by these systems, here we study the behavior of a
$d$-dimensional, long-range interacting, {\em neutral} plasma (in
contrast to a single component charged gas\cite{VladCoulomb}). The
system is described by a classical Hamiltonian,
\begin{equation}
  H = \oh \int_{\rv,\rv'} n(\rv) U_0(\rv - \rv')n({\rv'}) +E_c\sum_\alpha n_\alpha^2 \;,
  \label{Hlr} 
\end{equation}
where we have defined $\int_\rv \equiv \int d^dr$, with a "bare" long-range
power-law interaction
\beq U_0(\rv)= - K (r/a)^\omega\,,\;\; \omega=2-d+\sigma\;,
\label{Udef}
\eeq
where $a$ is a microscopic length scale.

We dub systems with $\sigma > 0$ ``super-Coulombic'' and $\sigma < 0$
``sub-Coulombic'', because such interactions are, respectively longer-
and shorter-ranged than a $d$-dimensional Coulombic interaction, Here
$K$, the constant interaction strength, satisfies $K\omega > 0$ (so as
to ensure attraction [repulsion] of opposite [like] charges), and in
equation (\ref{Hlr}) $E_c$ is the ``core'' energy, determined by
short-scale energetics. The charges $n(\br)$ are quantized in the
sense that
\beq n(\rv) = \sum_\alpha
n_\alpha\delta^d(\rv - \rv_\alpha)
 \label{quant}
 \eeq
 with integer charges $n_\alpha$.

 The Fourier transform of the interaction, computed in Appendix
 \ref{AppendixFT} is given by
 \beqn
U_0(\qv) = C(\sigma,d)K/q^{2+\sigma}  \,,
\label{UFT}
\eeqn
where $C(\sigma)$ is $O(a^{-\omega})$ function of $\sigma$ (and of
dimension of space $d$, argument that we have suppressed for
simplicity).  The precise value of $C(\sigma)$ is given in Appendix
\ref{AppendixFT}, and is unimportant. All that matters is that it is
finite for all $\sigma$ in the range of interest $-2<\sigma<2$, that
is, for interactions longer range than Coulombic, and the combination
of $C(\sigma) K$ is always positive, when the bare interaction between
opposite charges is attractive.

\subsection{Results}
\begin{figure}
    \centering
    \includegraphics[width=0.9\linewidth]{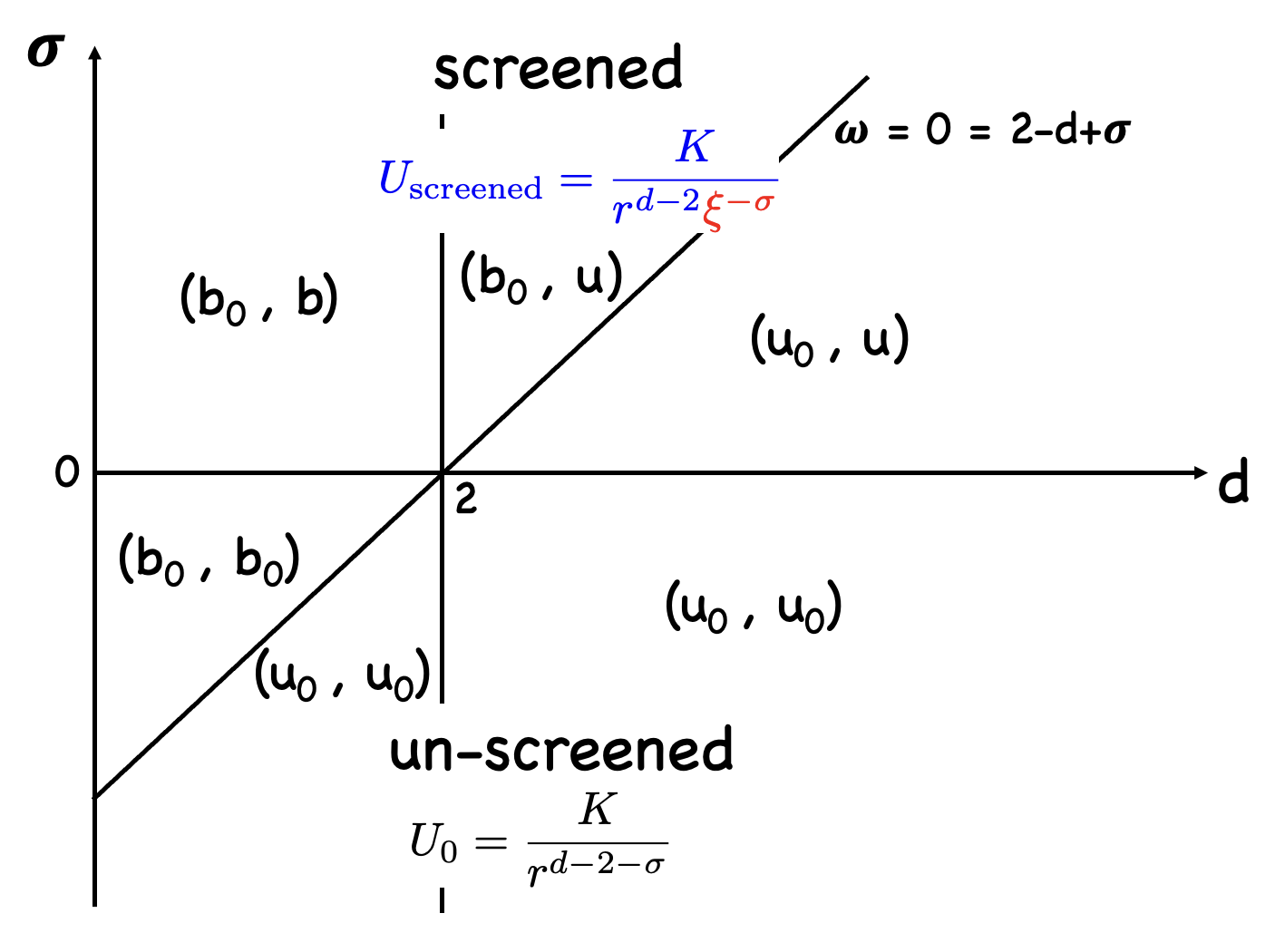}
   \caption{Illustration of the distinct screening regimes of a
     super/sub-Coulombic potential, which depend on the signs of $\sigma$
     and $\omega$, and on whether the spatial dimension $d$ is less
     than, greater than, or equal to $2$. Different regimes are
     labelled by the nature of the bare $U_0(r)$ (subscript $0$) and
     screened $U_{\text{eff}}(r)$ potentials, with $b$ and $u$
     denoting bound and unbound forms.}
    \label{regimesFig}
\end{figure}

Before turning to the analysis we first summarize our results. We find
that the power-law interacting plasma exhibits qualitatively different
behavior for $d>2$, $d=2$, and $d<2$, that also depends on the signs
of both $\sigma$ and $\omega$. As we will demonstrate, generically the
potentials can exhibit quite novel (crossover) screening of two
distinct types: (1) a power-law changing ``dielectric'' screening
beyond a length which we will call $\xi$, and (2) an unconventional
power-law ``Debye-Huckel'' screening beyond a length which we will
call $\xi_{DH}$.  The different behaviors and the regimes in which
they hold are summarized in Fig. \ref{regimesFig}.

\begin{figure}
    \centering
    \includegraphics[width=0.9\linewidth]{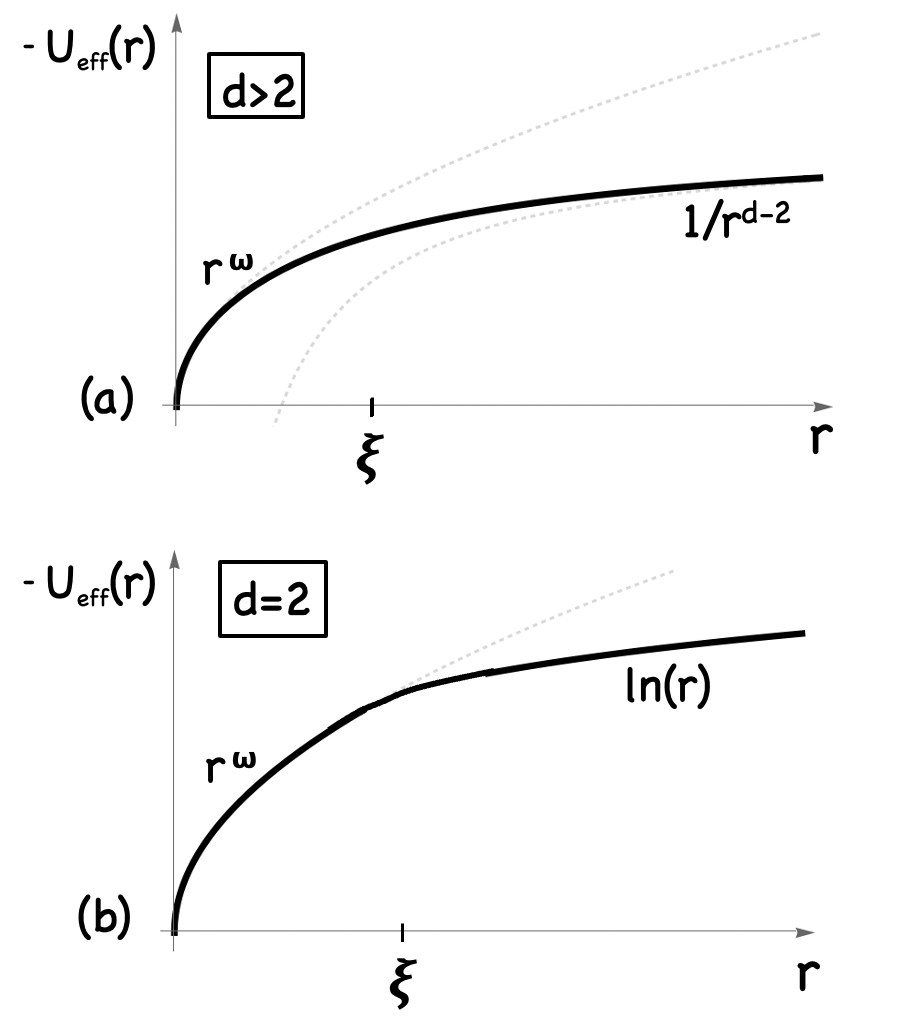}
    \caption{Illustration of the dielectric-like screening of a
      super-Coulomb potential for $\omega >0 $ down to a conventional
      Coulomb potential beyond the crossover scale $\xi$ given by
      equation \rf{xiResult}. (a) shows the case $d>2$ with a
      confining bare potential, $\omega > 0$ crossing over to the
      Coulomb non-confining potential, which approaches a non-zero
      constant (not indicated in the figure) as
      $-U_{\text{eff}}(r) \sim const. - 1/r^{d-2}$. Because such a
      decaying potential is non-binding, on scales longer than
      $\xi_{DH}$, the resulting conducting plasma will then undergo
      {\em unconventional} Debye-Huckel screening down to a more
      rapidly decaying {\em power-law} form \rf{unconventionalDH} (not
      illustrated in the figure).  In (b), the $d=2$ crossover beyond
      $\xi$ is to a logarithmically confining Coulomb form. As a
      result, opposite charges are bound into dipoles at low
      temperatures.  However, the system undergoes a conventional KT
      transition\cite{KT} as temperature is increased, with an
      enhanced transition temperature, to a phase exhibiting
      unconventional Debye-Huckel {\em power-law} interactions,
      (rather than the conventional exponentially decaying Yukawa
      potential), as discussed in the manuscript.}
    \label{Ueff}
\end{figure}

As illustrated there, for $\omega > 0$, at short scales the
two-component plasma (described by the microscopic potential
$U_0(r)\sim r^\omega$) behaves like a dielectric insulator with
charges tightly bound into neutral dipoles. We denote this behavior by
the symbol "$b_0$", which indicates that the bare potential is
"binding".  Then for $\sigma > 0$, we find that, as illustrated in
Fig.\ref{Ueff}, on intermediate length scales longer than the
crossover (screening) length
\beq
\xi=a\left[y^{-2}\left({K\over\kbt}\right)^{\left({d+2\over\omega}\right)-1}\right]^{1/\sigma}\times
O(1) \,,
\label{xiResult}
\eeq
(here $y\equiv e^{-\frac{E_c}{\kbt}}$ is the fugacity) a
super-Coulombic ($\sigma > 0$) potential is {\em universally}
``screened'' down to a conventional ($\sigma = 0$) Coulomb potential,
with a Fourier transform $U_{\text{eff}}(\bq)\sim 1/q^2$.  The
resulting effective potential $-U_{\text{eff}}(\br)$ between {\it
  opposite} charges in real space therefore crosses over from the
growing bare potential $-U_0(\br)$ given by (\ref{Udef}) to a
Coulombic, i.e., $U_{\text{eff}}(r)\propto r^{2-d}$ power-law. That is,
\bea U_0(\rv)&=& -K\left(\frac{r}{a}\right)^{2- d + \sigma}\;
\nonumber\\
&&\overset{r \gg \xi}{\underset{d\neq 2}{\xrightarrow{\hspace{20pt}}}}
\;U_{\text{eff}}(\br) = -K \left(\frac{\xi}{a}\right)^{\sigma}
\left(\frac{r}{a}\right)^{2-d}\,
\label{UcrossoverB}
\eea

As is clear from Fig. \ref{regimesFig}, what then happens to the
potential at asymptotically long scales depends on the spatial
dimension. For $d < 2$, the Coulomb power-law is "confining"; that is,
the potential between opposite charges goes to infinity as
$r\to\infty$, which makes the Boltzmann weight for widely separated
pairs vanish. Hence, oppositely charged pairs remain {\em bound}, even
after the effect of dielectric screening is taken into account. We
denote this by "b" (bound) in Fig. \ref{regimesFig}.  Because the
pairs remain tightly bound, this Coulombic power-law potential
persists out to arbitrarily large distances.  We emphasize that in
this case, the long-distance limit of the potential is still a power
law, albeit smaller than that of the bare potential, rather than an
exponential decay. That is, the "screening" on these intermediate
scales is more akin to the development of a non-zero dielectric
constant (albeit length scale dependent and equivalently one that
diverges as $q\to0$) in an {\em insulating dielectric} medium, than it
is to the familiar Debye-Huckel (DH) screening down to an
exponentially decaying Yukawa potential.

On the other hand, for $d > 2$, the Coulomb power-law that appears for
$r > \xi$ is no longer confining in the above-described sense, since
it vanishes as $r\to\infty$. As a result, at nonzero temperature, for
$r > \xi_{DH}$, where $\xi_{DH}>\xi$, the system will generically
undergo a second, "Debye-Huckel" screening of the Coulomb potential.
However, quite surprisingly, in contrast to a Coulombic conducting
plasma, here the Debye-Huckel screening is {\em unconventional},
leading to an asymptotically {\em power-law} (rather than decaying
exponential Yukawa) tail of the potential,
\begin{equation}
U_{\text{eff}}(\br)\approx \frac{C_{DH}}{r^{d+2+\sigma}} \,,
  \label{unconventionalDH}
\end{equation}
that decays faster than the Coulomb power-law
$U_{\text {Coulomb}}(\rv)\sim 1/r^{d-2}$ and (obviously) more slowly
than the conventional exponential Yukawa screening one finds for
Coulombic potentials.  The precise expression for $C_{DH}$ is given in
equation \rf{cdhdef} of section \rf{dh}, and \rf{Gsig} of Appendix
C. Its precise value depends on parameters of the model
($\sigma, K, E_c,\ldots$), as well as the spatial dimension $d$ and
$\kbt$.  However, the {\em sign} of $C_{DH}$ depends only on the
spatial dimension $d$ and the exponent $\sigma$. This sign is {\it
  negative} for all $\sigma$\Õs in the range $0<\sigma<1$ in $d=1$ and
$0<\sigma<2$ in $d=2$ and $d=3$. However, it becomes positive again
for $2<\sigma<3$ in $d=3$.

Recalling that $U_{\text{eff}}(\br)$ is the interaction between {\it
  like} charges, this implies that the cases enumerated above in which
$C_{DH}<0$, there is {\em overscreening}, i.e., an attractive
interaction tail between like charges.

Bare {\em confiining} ($\omega > 0$) {\it sub}-Coulombic
($\sigma < 0$) potentials, which are shorter-ranged than Coulombic,
remain {\em completely unscreened}
at low temperatures. However, rather surprisingly, we find that the
{\em non-confining} ($\omega < 0$) sub-Coulombic ($\sigma < 0$), bare
potentials do Debye-Huckel screen, but they do so far less effectively
than Coulombic potentials.  Specifically, they only screen down to a
{\em power-law} effective potential, identical to the DH power-law
screening form for $\sigma > 0$, $d>2$ given in
Eq.\rf{unconventionalDH}. Although such $\sigma < 0$, $\omega < 0$
potentials decay faster than their bare form, they decay more slowly
than both the exponential Yukawa screening one finds for Coulombic and
the power-law \rf{unconventionalDH} of super-Coulombic potentials
(since their $\sigma$ is negative).

Finally, as illustrated in Fig. \ref{Ueff}(b) for $d=2$ and
$\omega > 0$ and $\sigma> 0$, the screening is to a confining
logarithmic form.  Thus we strikingly predict a Kosterlitz-Thouless
(KT) phase transition\cite{KT} in a system with a
microscopically-power-law-confining potential (i.e., one with
$\omega > 0$).  This transition occurs at a temperature given in the
large core energy limit $E_c\gg\kbt$ by,
\beq k_B T_{KT}\approx{\sigma E_c\over2\ln\left({\sigma E_c\over
      K}\right)} \,.
\label{Tkt_results}
\eeq

We will demonstrate these results in two independent ways: (i) a
mapping to a sine-Gordon theory, and (ii) a dielectric theory. These
derivations are given in Sections \ref{secSG} and \ref{secDielectric},
respectively, with details of the analysis relegated to Appendices A
and B. Reassuringly, the two approaches give perfectly consistent
results.  In Sec. \ref{secKT} we will analyze the KT transition
temperature for the $d=2$ case. We will discuss the relation of the
two-component power-law interacting gas to long-range exchanged XY
model in Sec. \ref{secXY}, and conclude in Sec. \ref{secConclude}.

\vspace{0.51cm}

\section{Sine-Gordon screening}
\label{secSG}
We begin with a derivation of above results using a dual sine-Gordon
model description of the long-range interacting gas, defined by
Hamiltonian in (\ref{Hlr}). To this end, we put the model (\ref{Hlr})
on a lattice:
\begin{equation}
  H = \oh\sum_{\rv,\rv'} n_\rv U_0(\rv - \rv') n_{\rv'}+E_c\sum_\rv n^2_\rv\;,
  \label{Hlrlatt} 
\end{equation}
where the sums run over the sites $\rv$, $\rv'$ of a regular
lattice, and the $n_\rv$ are integers defined on those sites. The
partition function for this model at temperature $T$ is just the sum
of the Boltzmann weight for this Hamiltonian over all configurations
$\{n_\rv \}$ of the integers $n_\rv$ on the lattice:
\beq Z =
\sum_{\{n_\rv\}}e^{-H[\{n_\rv\}]/\kbt} \,.
\label{Z1}
\eeq
We now follow the standard treatment\cite{JKKN} of the ordinary
Coulomb gas by introducing a Hubbard-Stratonovich field $\theta_\rv$
to mediate the long-range interaction. That is, we rewrite the
partition function (\ref{Z1}) as
\bew
\beq
Z=\sum_{\{n_\rv\}}\prod_\rv\left[\int d\theta_\rv\right]
\exp\bigg[- {\kbt\over2C(\sigma)K}\int_{\qv}\; q^{2
  +\sigma}|\theta_{\qv}|^2-\sum_\rv\bigg(\frac{E_c}{\kbt} n^2_\rv-i\theta_\rv n_\rv\bigg)\bigg] \,,
\label{Z2}
\eeq
\ew
where $\theta_\bq$ is simply the Fourier transform of $\theta_\rv$.
Performing the Gaussian integral over $\theta_\bq$ in this expression
is readily seen to recover the interaction \rf{Hlrlatt}
between charges in Fourier space of the original model.

Reorganizing the sums and products in (\ref{Z2}) gives

\beq
Z=\prod_\rv \left[\int d\theta_\rv\right]
e^{-\frac{1}{\kbt}\sum_\rv V[\theta_\rv]
-\frac{\tilde K}{2\kbt}\int_{\qv}\; q^{2+\sigma}|\theta_{\qv}|^2}\, , 
\label{Z3}
\eeq
where the summation over {\em integer} charges $n_\rv$ on each lattice site $\rv$ 
gives the  {\it periodic} "Villain potential"\cite{Villain} of $\theta_\rv$,
\beq
V[\theta_\rv]\equiv -\kbt\ln\bigg[\sum_{n_\rv=-\infty}^\infty
e^{-\frac{E_c}{\kbt} n^2_\rv+i\theta_\rv n_\rv}\bigg] \,,
\label{villaindef}
\eeq
and we have defined the "dual stiffness"
\beq
\tK \equiv {(\kbt)^2\over C(\sigma)K} \,.
\label{tKdef}
\eeq
Note that $V[\theta_\rv]$ is manifestly a periodic function of
$\theta_\rv$ with period $2\pi$.
   
Going back to the continuum, the partition function (\ref{Z3}) is then
given by a path integral
($\int[d\theta(\rv)]\equiv \prod_\rv[\int d\theta_\rv]$) over the field
$\theta(\rv)$,
$Z= \int[d\theta(\rv)] e^{-\tilde H[\theta(\rv)]/\kbt}$,
  with the generalized dual {\em long-range} sine-Gordon model
  Hamiltonian, given by
\begin{eqnarray}
  \tilde  H   &=& \oh \tilde K\int_{\qv}\; q^{2+\sigma}|\theta_{\qv}|^2
- \sum_{n=1}^\infty g_n\int_\xv\cos(n\theta(\xv))\nn,\\
\label{sgHlr}
\end{eqnarray}
where the $g_n$'s are $a^d$ times the expansion coefficients in the
Fourier series for the periodic potential $-V[\theta(\rv)]$, with
$a^d$ the hypervolume of the unit cell of the lattice we introduced
earlier.  Because higher harmonics are always less relevant, and much
smaller\cite{JKKN}, we will focus on the first one, $g_1$.

Indeed, it is straightforward to see that, for small "fugacity"
$y\equiv e^{-\frac{E_c}{\kbt}}\ll1$, the sum in \rf{villaindef} is
dominated by the $n_\br=0\,,\pm1$ terms giving
\beq
V[\theta_\rv]\approx -\kbt\ln\bigg[1+2y\cos(\theta_\rv) \bigg] \approx
-2\kbt y\cos(\theta_\br)\,,
\label{villainapprox}
\eeq 
from which we can read off \beq g_1\approx2\kbt ya^{-d}=2\kbt
e^{-\frac{E_c}{\kbt}}a^{-d} \,,
\label{g1est}
\eeq
where the factor of the inverse of the unit cell volume $a^d$ arises,
as mentioned earlier, from going over from the lattice to the
continuum.  By extending this argument to terms higher order in the
fugacity $y$, one can show that, in the limit of small fugacity, the
higher harmonics
$g_{n>1}\propto y^{n}\propto e^{-\frac{n E_c}{\kbt}}\ll g_1$, as
asserted earlier.

One might naively expect that the behavior of the model in two
dimensions ($2D$) would differ significantly and qualitatively from
that of the conventional-gradient-elasticity ($\sigma = 0$)
sine-Gordon model.  For $\sigma = 0$ in $2D$, the cosine terms are
"relevant", in the RG sense of changing the long-wavelength behavior
of the system, for large stiffness $\tilde K$, which corresponds to
high temperatures of the original model \rf{Hlr}, and to the unbinding
of charge dipoles\cite{JKKN}. In contrast, for our system, the
``soft'' $q^{2+\sigma}$ elasticity of the {\em long-range} sine-Gordon
model \rf{sgHlr} would naively appear to ensure that for $\sigma > 0$,
the cosine nonlinearities are {\it always} -- for {\it all} $T$ and
$K$ -- {\em irrelevant} for weak $g_n$ couplings. That is, the
enhanced $\theta(\rv)$ fluctuations at long wavelengths always
``average'' away the weak cosine nonlinearities.  In the context of a
2D XY model, this is simply a restatement in the dual language of the
naive intuition that vortices would {\it always} be bound by a
potential that is longer ranged than logarithmic.

However, our key and surprising discovery is that this naive
conclusion is in fact incorrect. It is invalidated by universal
screening of the long-range interaction $1/q^{2+\sigma}$ in \rf{Hlr},
which reduces that interaction to the conventional Coulomb $1/q^2$
interaction, for all $\sigma>0$.

This is easiest to see directly in the long-range sine-Gordon
formulation \rf{sgHlr} through a renormalization group (RG) analysis
done perturbatively in the $g_n$'s.  As we detail in the Appendix A,
the renormalization group always generates a non-zero "spin-wave
stiffness", $\oh\kappa (\grad\theta)^2$ term in the sine-Gordon model,
where $\kappa = O(g^2)$.  At sufficiently long scales $r$ --
specifically for $r \gg \xi \sim g^{-2/\sigma}$, (a detailed
expression for $\xi$ is given in \rf{xiSG}, and derived in the SM) --
for the super-Coulombic case $\sigma > 0$, this dominates over the
``soft'' long-range ``elasticity'' \rf{sgHlr} at small $q$, since
$\tilde K q^{2+\sigma} \ll \kappa q^2$ for $q\xi\ll 1$.  Reversing the
above duality procedure, one can see that this gives an effective
interaction between charges at these longer length scales that is the
Fourier transform of $1/q^2$; i.e., the conventional Coulomb
interaction $1/r^{d-2}$ in $d>2$, and $\ln{r}$ in $d=2$, as claimed in
Eq. \rf{UcrossoverB}.  This Coulombic interaction replaces the longer
ranged $U_0(\rv)$ in \rf{Hlr} for scales longer than the crossover
length $\xi$.  In the Appendix A, we do this renormalization group
calculation of both $\kappa$ and the crossover length $\xi$ in detail,
and show that at long scales, the effective ``spin-wave'' stiffness
$\kappa$ is given by
\bea \kappa &=&\kbt \left({\kbt\over
    K}\right)^{(d+2)/\omega} a^{2-d}e^{-2E_c/\kbt} \times O(1)\,, \nn\;\;\;\;\;\;\\
&=&\frac{(\kbt)^2 a^\omega}{K\xi^\sigma} \times O(1) \,,
\label{kappaRG}
\eea
with the crossover length
\beq
\xi=a\left[\left({K\over\kbt}\right)^{\left({d+2\over\omega}\right)-1}\left({1\over
      y^2}\right)\right]^{1/\sigma}\times O(1) \,.
\label{xiSG}
\eeq
Undoing the duality procedure then implies that the effective
screened interaction between charges at length scales much larger than
the crossover length scale $\xi$ is given by the conventional Coulomb
one,
\beq U_{\rm eff}(\br)=\frac{K_{\rm Coulomb}}{r^{d-2}} \,,
\label{coulomb}
\eeq
with the effective Coulombic coupling strength $K_{\rm Coulomb}$ given by
\beq
K_{\rm Coulomb}={(\kbt)^2\over\kappa}= Ka^{d-2}(\xi/a)^\sigma\times O(1) \,.
\label{Kcoul}
\eeq
%

In contrast, for a bare potential that is shorter-ranged than
Coulombic ($\sigma < 0$), it is clear by inspecting the Hamiltonian
(\ref{sgHlr}) that the $q^{2+\sigma}$ term will {\it always} dominate
over the fluctuation generated $q^2$ term. As a result, the effective
interaction $U_{\text{eff}}(r)$ between vortices retains its bare
$r^\omega$ form.

\section{Dielectric screening}
\label{secDielectric}
A more familiar, although algebraically
slightly more complicated argument for the above universal screening
prediction is a complementary dielectric medium analysis of the
long-range interaction \rf{Hlr}.

This calculation is based on the observation that even tightly bounds
pairs of charges create a "dipole" field at large distances. These
tightly bound dipoles tend to align with an externally imposed field,
in such a way as to cancel off the external field. This is the
mechanism that generates a non-unit dielectric constant in a
dielectric insulator, as opposed to the Debye-Huckel screening down to
a Yukawa potential that takes place in a conducting plasma of free
charges.
 
As a result of these "induced" dipoles, the effective interaction
between two unit test charges introduced to our system is
given by:
\beq
U_{\text{eff}}(\br)=U_0(\br)+U_i(\br) \,,
\label{screened1}
\eeq
where the induced dipole potential at the point $\br$ is given by
\beq U_i(\br)=\sum_\alpha
U_{d\alpha}(\br-\br_\alpha) \,,
\label{induced1}
\eeq
where $U_{d\alpha}(\br)$ is the potential induced at point $\br$ by a
dipole at $\br_\alpha$.   For a pair
of equal and opposite charges $n_\alpha$ (screening dipole
constituents) separated from each other by $\bv_\alpha$, this is given
by
\bew
\beq
 U_{d\alpha}(\br-\br_\alpha)=n_\alpha[U_0(\br-\br_\alpha)-U_0(\br-\br_\alpha+\bv_\alpha)]\approx-n_\alpha\bv_\alpha\cdot\grad U_0(\br-\br_\alpha)
 \equiv-\bm{\pv}_\alpha\cdot\grad U_0(\br-\br_\alpha)
 \label{dipole1}
\eeq
\ew
where $\bm{\pv}_\alpha\equiv n_\alpha\bv_\alpha$ is the dipole moment
of the $\alpha$'th dipole. 
The induced potential $U_i(\br)$ is thus given by
\beq
U_i(\br)=-\sum_\alpha \bm{\pv}_\alpha\cdot\grad U_0(\br-\br_\alpha) =-\int_{\rv'} \bp(\br') \cdot\grad U_0(\br-\br')
\,,
\label{induced2}
\eeq
where 
\beq
\bp(\br')\equiv\sum_\alpha \bm{\pv}_\alpha\delta^d(\br'-\br_\alpha)
\label{dipdensdef}
\eeq is the local vectorial dipole density at $\br'$.  Using this
result in our expression \rf{induced2} for the dipole potential, and
putting the result in equation \rf{screened1} leads to the total
screened potential
\beq
U_{\text{eff}}(\br)=U_0(\br)-\int_{\rv'} \bp(\br') \cdot\grad U_0(\br-\br') \,.
\label{screened2}
\eeq


As in ordinary electrostatic dielectric theory, within linear
response, the local dipole density $\bp(\br')$ is proportional to the
local "electric field":
\beq \bp(\br)=-\chi\grad U_{\text{eff}}(\br) \,,
\label{susc}
\eeq
where $\chi$ is the susceptibility.  We calculate $\chi$ using the
Boltzmann statistics of an isolated dipole in an external field in the
Appendix B\cite{SM}, with the result over a wide range of
parameters\cite{range} given by
\beqn \chi={y^2\over \kbt
  a^{d-2}}\left({\kbt\over
    K}\right)^{\left({d+2\over\omega}\right)}\times O(1)\,.
\label{chiMT}
\eeqn
The only feature of this result that we really need in the current
discussion is that $\chi$ is non-zero, and finite.  Using the linear
response relation (\ref{susc}) inside (\ref{screened2}) gives,
\beq
U_{\text{eff}}(\br)=U_0(\br) + \chi\int_{\rv'} \grad' U_{\text{eff}}(\br') \cdot\grad
U_0(\br-\br') \,,
\label{sc}
\eeq
which can be straightforwardly solved in Fourier space, giving 
\beq
U_{\text{eff}}(\bq)=\frac{U_0(\bq)}{1+\chi q^2 U_0(\bq)}\,.
\label{ussolv}
\eeq

%
Now, since, for $\sigma > 0$, $q^2 U_0(\bq)= K a^{-\omega}/q^{\sigma}$
diverges as $q\to 0$, at sufficiently small $q$ the term
$\chi q^2 U_0(\bq)$ in the denominator of (\ref{ussolv}) dominates
over the $1$. More precisely, we see that this will occur for
$q\ll\xi^{-1}$, where the screening length is given by
\beq
\xi=a^{\omega\over\sigma}\left(K\chi\right)^{-{1\over\sigma}}=a\left[\left({K\over\kbt}\right)^{\left({d+2\over\omega}\right)-1}\left({1\over
      y^2}\right)\right]^{1\over\sigma}\times O(1) \,,
\label{xidielectric2}
\eeq
as advertised in \rf{xiResult} of the Introduction.  Note from the
second equality that this is precisely the same crossover length we
found in the sine-Gordon approach, which is a non-trivial and
reassuring check on the validity of both the sine-Gordon and this
dielectric approach.
%
Thus for $q\ll\xi^{-1}$ we obtain,
\beq U_{\text{eff}}(\bq)\approx \frac{1}{\chi
  q^2}\equiv\frac{K_{\rm Coulomb}}{q^2}\;,\;\;\mbox{for
  $q\ll \xi^{-1}\equiv a^{-{\omega\over\sigma}}(\chi
  K)^{1/\sigma}$}\,,
\label{ussolvasym}
\eeq
where 
\beq
K_{\rm Coulomb}=\chi^{-1}=Ka^{d-2}(\xi/a)^\sigma \times O(1)\,,
\label{Kcoul2}
\eeq
Reassuringly, the sine-Gordon and the dielectric analyses agree in
predicting that, quite generically, the screened potential
asymptotically crosses over to the conventional Coulomb potential for
{\it any} bare potential that is longer ranged than Coulombic
($\sigma > 0$), in {\it any} spatial dimension.  Both calculations
also give the same strength of that effective Coulomb potential.
  
Note also that if the bare interaction is itself Coulombic -- that is,
if $\sigma = 0$, the denominator in \rf{ussolv} is a finite constant
larger than $1$, and our result simply reduces to a Coulomb
interaction reduced by the conventional dielectric constant of the
medium. For a bare potential that is shorter-ranged than Coulombic
($\sigma < 0$), we also recover the result of the above in the
sine-Gordon analysis, since, at small $\bq$, the $\chi q^2 U_0(\bq)$
term in the denominator of (\ref{ussolv}) is $\ll1$ (since in this
case $U_0(\bq)\ll1/q^2$ as $\bq\to{\bf 0}$). As a result, at
sufficiently small $\bq$ and large $\br$, the effective potential
reduces to the bare potential.

\section{Implications for the various interaction types}
As we now detail, the above results lead to a number of important
implications for different cases of $\omega$, $\sigma$ and spatial
dimension $d$, leading to two distinct ``power-law dielectric'' and
``power-law Debye-Huckel'' screening types, as summarized in
Fig.\ref{regimesFig}.

\subsection{Screening of non-confining ($\omega < 0$) {\bf
    sub}-Coulombic ($ -2 < \sigma < 0$) potentials}\label{dh}
We first observe that {\em sub}-Coulombic potentials (i.e., those with
$\sigma<0$) do not exhibit the dielectric-like screening we studied in
Secs. \ref{secSG} and \ref{secDielectric} above. This can be seen in
the sine-Gordon approach from the fact that the generated short-range
$\kappa$ elasticity is subdominant to the long-range $\tilde K$
elasticity, and from the dielectric analysis based on Eq.\rf{ussolv},
as discussed above.

However, for the non-confining case of $\omega < 0$, such {\em
  sub}-Coulombic $\sigma<0$ potentials {\it do} in fact exhibit the
analog of Debye-Huckel screening, but with a quite surprising {\em
  power-law} (rather than exponential) screened form. This can be
derived from the sine-Gordon theory developed above by noting that,
for $\omega < 0$, the cosine terms are {\it always} relevant, with
eigenvalue $d$, which implies deconfinement of charges.  This means
that, at sufficiently long distances,
$r \gg \xi^-_{DH} = (\tilde K/g)^{1/(2+\sigma)}$ (the "-" superscript
is to distinguish this length from the analogous length $\xi_{DH}$ for
$\sigma > 0$, which we discuss below), the $-g\int_\rv \cos\theta_\rv$
always reduces to a non-zero effective ``mass'' $g$ for the dual field
$\theta_\rv$ in \rf{sgHlr}. That is, the effective long-wavelength
model becomes
\begin{eqnarray}
  \tilde  H   &=& \oh \int_{\qv}\; (g+\tilde Kq^{2+\sigma})|\theta_{\qv}|^2 \,,
\label{sig<0H}
\end{eqnarray}
which in turn implies that the $\theta_\rv$ correlations in Fourier
space are given by
\beq
\langle|\theta_\bq|^2\rangle={\kbt\over g+\tilde Kq^{2+\sigma}} \,.
\label{sgcorrsig<0}
\eeq
By reversing our dual mapping onto the sine-Gordon theory, this
implies that the Fourier transformed effective screened potential on
wavenumber scales smaller than $(\xi^-_{DH})^{-1}$ is given by
\beq 
U_{\text{eff}}(\bq)= {(\kbt)^2 \over g+ \tilde Kq^{2+\sigma}}
\,.
\label{Usub1}
\eeq
%
%
If we expand $U_{\text{eff}}(\bq)$ for small $\bq$, the first
non-analytic term is
$-\left({(\kbt)^2\tilde K\over g^2}\right)q^{2+\sigma}$, which, as we
show in Appendix C, implies
\beqn
U_{\text{eff}}(\br)&\approx&\frac{C_{DH}}{r^{d+2+\sigma}} \,,
\label{Usubfin}
\eeqn
with
\beq
C_{DH}=G(\sigma,d)\left({(\kbt)^2\tilde K\over g^2}\right) \,,
\label{cdhdef}
\eeq
where the dimensionless constant $G(\sigma,d)$ depends only on the
spatial dimension $d$ and the exponent $\sigma$. The precise, rather
complicated expression for $G(\sigma,d)$ is given by equation
\rf{Gsig} of Appendix C. All we need to know here is that
$G(\sigma,d)$ is finite and {\it positive} for all $\sigma$ in the
range $-2<\sigma<0$, and vanishes at $\sigma=0$. This vanishing as
$\sigma\to0$ implies that there is no long-ranged power-law tail for
Coulombic potentials, in agreement with the well-known result that
Coulombic potentials are Debye-Huckel screened down to exponentially
decaying Yukawa forms.

Our result \rf{Usubfin} implies, rather surprisingly, that bare {\it
  sub}-Coulombic potentials with non-even integer $\sigma < 0$, which
are shorter-ranged than Coulombic potentials, screen {\it less}
effectively than those potentials on scales longer than $\xi^-_{DH}$,
leading to effective {\em power-law} screened potentials at the
longest distances. Obviously, such potentials are {\it longer}-ranged
than the conventional exponentially screened Yukawa potential.

This result, taken together with our above result \rf{Usubfin} for
$-2 < \sigma < 0$, implies that the longest ranged effective potential
(in the sense of decaying with the smallest power law with distance)
occurs when $\sigma=-2$, in which case $U_{\rm eff}\propto r^{-d}$.

\subsection{Absence of screening for $\sigma<-2$}

For $\sigma<-2$, the bare potential \rf{Hlr} falls into the class of
potentials considered in part \rf{dhapp} of Appendix C, that is, its
volume integral is finite, and, hence, its Fourier transform is finite
as $\bq\to{\bf 0}$.  More precisely, using equation \rf{g7} of part
\rf{dhapp} of Appendix C with $\gamma=d-2-\sigma$, we see that the
Fourier transform $U(\bq)$ of the bare potential can be written
\bea
U(\bq) &=& U_{\rm a}(\bq)+A(\sigma, d)
K a^{d-2-\sigma} q^{-(2+\sigma)}\;,\label{sig<-2.1}
\eea
where $U_{\rm a}(\bq)$ is an analytic function of $\bq$ which does
not vanish as $\bq\to0$, and which depends only on $q^2$, and
$A(\sigma, d)$ is an unimportant constant. Note that for $\sigma<-2$
that we are considering here, the exponent $-(2+\sigma)$ of the
explicitly displayed power of $q$ in \rf{sig<-2.1} is positive. Hence,
this term is sub-dominant relative to the $U_{\rm a}(\bq)$ term as
$\bq\to0$ since that term does not vanish in that limit.

Now, when we perform the duality transformation on the plasma model
\rf{Hlr}, we obtain the sine-Gordon model Hamiltonian
\begin{eqnarray}
  \tilde  H   &=& \oh \int_{\qv}\;  G(\bq)|\theta_{\qv}|^2
- \sum_{n=1}^\infty g_n\int_\xv\cos(n\theta(\xv))\,,
\label{sgHlrsig<-2}
\end{eqnarray}
where we have defined
\beq
G(\bq)\equiv\left({(\kbt)^2\over U_{\rm a}(\bq)+A(\sigma, d)
K a^{d-2-\sigma} q^{-(2+\sigma)}}\right) \,.
\label{Gdef}
\eeq

Since the $q^{-(2+\sigma)}$ term in this expression is subdominant
relative to the $U_{\rm a}(\bq)$ term as $\bq\to0$, if we expand
$G(\bq)$ for small $\bq$, we get
\bea
G(\bq) &=& G_{\rm a}(\bq)+B(\sigma, d)
 q^{-(2+\sigma)}\,,\label{Gexp}
 \eea
 where $G_{\rm a}(\bq)$ is an analytic function of $\bq$ which does
 not vanish as $\bq\to0$, and which depends only on $q^2$, and
 $B(\sigma, d)$ is another unimportant constant.

Now consider the effect of renormalizing this model. Unlike the
$\sigma>-2$ cases, the analytic structure of $G(\bq)$ will not change
upon renormalization: it already has a non-zero ``mass'', so the mass
generated by the cosine in \rf{sgHlrsig<-2} only adds to something
that is already present. Likewise, generated $q^2$ terms will only add
finite renormalizations to the $q^2$ terms already present in
\rf{Gexp}.

Hence, the analytic structure of the full sine-Gordon model
\rf{sgHlrsig<-2} does not change upon renormalization.  Thus, if we
renormalize, and then undo our duality to obtain the effective
potential, we will get an effective potential of {\it exactly} the
same analytic structure as the bare potential. That is, it will still
fall off like $r^\omega$ with $\omega=2-d+\sigma$. Namely, there is
{\it no} screening in this case at all. Rather, the effect of the
other charges in the system on the interaction of two test charges is
more like the finite dielectric constant that occurs in Coulombic
insulators: the {\it coefficient} of the long ranged tail of the
interaction is changed by those other charges, but the power law of
the interaction is not.

This result, taken together with our above result \rf{Usubfin} for
$-2 < \sigma < 0$, implies that the longest ranged effective potential
(in the sense of decaying with the smallest power law with distance)
occurs when $\sigma=-2$, in which case $U_{\rm eff}(r)\propto r^{-d}$.

\subsection{Screening for $d>2$ super-Coulombic ($\sigma > 0$)
  potentials}

In $d>2$ the effect of the power-law dielectric-like screening we have
derived in Secs. \ref{secSG} and \ref{secDielectric} is particularly
striking for a bare potential that is {\em confining} on short scales,
that is, one with $\omega > 0$.  In this case, naively (i.e., ignoring
the dielectric screening that we predict beyond $\xi$), one would
expect that the charges would {\it always} be confined.  However, in
fact, as demonstrated above, on length scales beyond the crossover
scale $\xi$ the effective potential is screened down to a Coulomb
potential. Since this potential {\it vanishes} as $\br\to\infty$ for
$d>2$, there can be no confinement of charges. Hence, we predict that
the exact opposite of the naive expectation is true: charges are {\it
  never} confined, as is clear from Fig.~(\ref{Ueff}b). Hence, there
is only a single, deconfined phase in such a super-Coulombic neutral
plasma in $d>2$.

Furthermore, once the bare confining super-Coulombic potential
($\omega > 0$, $\sigma > 0$) screens down to a Coulomb potential,
which is non-confining for $d > 2$, the two-component gas is a
conducting plasma that is subject to a Debye-Huckel screening.  In
terms of the sine-Gordon model, this corresponds to the observation
that for $d > 2$ the $-g\int_\rv \cos(\theta_\rv)$ is always relevant
(charges are always deconfined), reducing to a ``massive'' $\theta$
model,
\begin{eqnarray}
  \tilde  H^{\gg\xi}_{DH}   &=& \oh\int_{\qv}\;\left[\tilde K q^{2+\sigma} + \kappa q^2 + g_R \right]|\theta_{\qv}|^2 \,,
\label{sgDH}
\end{eqnarray}
with the effective coupling $g_R\equiv g e^{-c\xi^\omega}$ reduced by
coarse-graining out to scales $\xi$ (as detailed in Appendix A). Here
$c$ is a positive constant of order $a^{-\omega}$.

Thus, beyond the Debye-Huckel screening length
$\xi_{DH}=(\kappa e^{c\xi^\omega}/g)^{1/2}$, the effective potential
exhibits further deconfined plasma screening.  One may naively think
that it is possible to simply drop the $\tilde K q^{2+\sigma}$ term in
\rf{sgDH}, since for $\sigma>0$ it is higher order in $q$ than the
generated $\kappa q^2$ term.  {\em If} it were possible to do so, one
would then obtain the conventional exponentially short-ranged
potential of the Yukawa type.  However, as we will see below, beyond
$\xi_{DH}$, the resulting real-space Debye-Huckel screened potential
is generically (other than for an even integer $\sigma$) quite
unconventional {\em power-law } in $r$, as advertised in
Eq.\rf{unconventionalDH} of the Introduction, rather than the decaying
exponential Yukawa potential that one obtains for a conventional
Coulomb plasma.

To see this somewhat surprising result, we first note that \rf{sgDH}
gives,
\beq \langle|\theta_\bq|^2\rangle={\kbt\over g_R+\kappa q^2 +\tilde
  Kq^{2+\sigma}} \,,
\label{sgcorrsig>0}
\eeq
that, by reversing our dual mapping onto the sine-Gordon theory,
implies that the Fourier transformed effective screened potential
$U_{\text{eff}}(\bq)$ on wavevectors smaller than
$\xi_{DH}^{-1}$ is given by
\beq U_{\text{eff}}(\bq)= {(\kbt)^2\over g_R+\kappa q^2 + \tilde Kq^{2+\sigma}}
\,.
\label{Usup1}
\eeq

If we expand $U_{\text{eff}}(\bq)$ for small $\bq$, the first
non-analytic term is
$-\left({(\kbt)^2\tilde K\over g_R^2}\right)q^{2+\sigma}$, which, as
we show in Appendix C, implies
\beqn U_{\text{eff}}(\br)&\approx&\frac{C^R_{DH}}{r^{d+2+\sigma}} \,,
\label{Usupfin}
\eeqn
with
\beq
C^R_{DH}=G(\sigma,d)\left({(\kbt)^2\tilde K\over g_R^2}\right) \,,
\label{cdhdefR}
\eeq
where the dimensionless constant $G(\sigma,d)$ depends only on the
spatial dimension $d$ and the exponent $\sigma$, and has the same form
\rf{Gsig} as for the sub-Colombic case, but now with $\sigma>0$. In
$d=3$ (which is obviously the only physically relevant case with
$d>2$), $G(\sigma,d)$, as can be seen from the plot Fig. \rf{gplot3},
is positive for $2<\sigma$, but {\it negative} for $0<\sigma<2$.
Recalling that $U_{\text{eff}}(\br)$ is defined as the interaction
between {\it like} charges, we see that a positive $G(\sigma,d)$
implies an attractive interaction between opposite charges (like the
bare interaction in \rf{Udef}), while a {\it negative} $G(\sigma,d)$
implies a {\it repulsive} interaction between opposite charges. We
call this phenomenon ``overscreening''.

This overscreening only occurs when $\sigma$ lies in the range
$0<\sigma<2$, which corresponds in $d=3$ to power-law potentials that
fall off more slowly than a Coulombic $1/r$ potential (or grow), but
do not grow as rapidly as a linear potential. That is, the power-law
$\omega$ in \rf{Udef} lies in the range $-1<\omega<1$.

The vanishing of $G(\sigma,d)$ as $\sigma\to0$ implies that there is
no long-ranged power-law tail for Coulombic potentials, in agreement
with the well-known result that Coulombic potentials are Debye-Huckel
screened down to exponentially decaying Yukawa forms.

The vanishing of $G(\sigma,d)$ as $\sigma\to2$ in $d=3$ is simply a
consequence of the fact that the inverse Fourier transform of the bare
potential is analytic in this case. It implies that there is no
long-ranged power law tail to the effective interaction at long
distances in this case. Instead, in this case, like the Coulomb case,
we will have an exponentially decaying Yukawa potential at the longest
distances.

\subsection{Competition between dielectric and Debye-Huckel crossovers}
\label{twoCrossovers}
Above we have discussed two qualitatively distinct types of screening:
``dielectric-like'' (discussed so far) screening that converts
super-Coulombic potentials to Coulombic ones, and (unconventional)
Debye-Huckel screening. Interestingly, both types of screening lead to
asymptotical power-law potentials.

To summarize, we have considered four distinct cases:

\noindent(i) for $\sigma <0$, the dielectric screening is absent and
only power-law Debye-Huckel screening takes place for $\omega < 0$;

\noindent(ii) for $\sigma > 0$ and $d \leq 2$ (and concommitantly
$\omega > 0$), at low temperatures the dielectric screening always
takes place first, since the gas is confining below a critical
temperature, and Debye-Huckel screening only takes place above the
deconfining phase transition temperature.

\noindent(iii) for $\sigma > 0$ and $d > 2$, both dielectric and
Debye-Huckel screening take place. However, for $\omega > 0$,
$-g\cos\theta$ is irrelevant on scales shorter than dielectric
screening length $\xi$. As a result, the Debye-Huckel screening always
takes place on scales {\em longer} than $\xi$, with $\xi_{DH}$ always
longer than $\xi$;

\noindent(iv) in contrast to (iii), for $\sigma > 0$ and $d > 2$, but
$\omega < 0$, $-g\cos\theta$ is relevant and both dielectric and
Debye-Huckel screening can take place.

The last case (iv) exhibits a competition between the two types of
screening, determined by the relative size of the corresponding
screening lengths, $\xi$ and $\xi_{DH}$. For (a) $\xi\ll \xi_{DH}$ the
intermediate dielectric screening regime will survive up to length
$\xi$, followed by Debye-Huckel screening beyond scale $\xi_{DH}$.
Alternatively, for (b) $\xi\gg \xi_{DH}$, $-g\cos\theta$ is relevant
and leads to a ``mass'' for $\theta$ on scales beyond $\xi_{DH}$. This
thereby precludes the intermediate dielectric screening regime.

To determine the range of parameters for the regimes (iv) (a) and (iv)
(b), we examine the ratio

\bea
\rho& \equiv& \xi/\xi_{DH}\nn\\
&=&a\left[e^{\frac{2E_c}{\kbt}}\left({K\over\kbt}\right)^{\left({d+2\over\omega}\right)-1}\right]^{1/\sigma}
\times (2\kbt e^{-\frac{E_c}{\kbt}}a^{-d}/\tilde K)^{1/(2+\sigma)}\nn\\
 &=&e^{\frac{\zeta_1E_c}{\kbt}}\left({K\over\kbt}\right)^{\zeta_2}\,,
\label{rho}
\eea
where in the final equality, we have used equation \rf{tKdef} to
rewrite $\tK$ in terms of the original interaction strength $K$ of the
bare potential, and ignored factors of $O(1)$.  We have also in the
final equality defined
\beq
\zeta_1\equiv{4+\sigma\over\sigma(2+\sigma)} \sep
\zeta_2\equiv{2(1+\sigma)\over\sigma(2+\sigma)} \,.
\label{zetas}
\eeq
Note that both $\zeta_{1,2}>0$, since we are considering $\sigma>0$ here.

Clearly, for low temperatures, specifically $\kbt\ll E_c$, the ratio
$\rho$ gets very large, and, hence, there is no intermediate
dielectric regime. Instead, one crosses over directly from the bare
interaction to the anomalous (power-law) Debye-Huckel screened one.

As we raise the temperature into the regime $\kbt\gtrsim E_c$ and
$\kbt\gg K$, the ratio $\rho\ll1$, and we will have an intermediate
dielectric regime. To say this in another way, with increasing
distance $r$, the interaction will first cross over from the bare
interaction to a $d$-dimensional Coulomb interaction $r^{2-d}$, and
then, at a much larger length scale $\xi_{DH}$, cross over to the
unconventional (power-law) Debye-Huckel screened interaction
\rf{unconventionalDH}.

\subsection{Kosterlitz-Thouless transition in $d=2$}
\label{secKT}
The other important implication of the screening predicted above is in
two dimensions. For a gas interacting with a bare confining potential,
(i.e., $\omega > 0$), one might naively again expect that charges are
always bound.  In contrast, what we actually predict is screening to a
logarithmic potential, as illustrated in Fig.(\ref{Ueff}a), beyond the
crossover scale $\xi$ in Eq.\rf{xiResult}.  This implies that such a
gas will therefore undergo a conventional Kosterlitz-Thouless
unbinding phase transition at
$k_BT_{KT}/K_{\rm Coulomb}(T_{KT}) = 1/4~$\cite{KT}. Utilizing our
expression \rf{Kcoul2} for $K_{\rm Coulomb}$, and setting $d=2$ (i.e.,
$\omega=\sigma$), we see that this implies the KT transition
temperature for a super-Coulomb gas obeys 
\beqn k_B T_{KT} &=& K
\left({\xi(T_{KT})\over a}\right)^\sigma\times O(1)\;,\\
&=& {k_BT_{KT}\over y^2}
\left({K\over\kbt_{KT}}\right)^{4\over\sigma}\times O(1)\nn\\
&=& {k_BT_{KT}\over4}
\left({K\over\kbt_{KT}}\right)^{4\over\sigma}e^{2
  E_c/k_B T_{KT}}\times O(1)\,.\nonumber\\
\label{Tkt}
\eeqn
The solution of this equation for $T_{KT}$ is
\beq
 k_B T_{KT}={\sigma E_c\over2W_0\left({\sigma E_c\over K}\times O(1)\right)}\;,
 \label{Tkt sol}
 \eeq
 where $W_0(u)$ is the first branch of the Lambert W
 function\cite{Lamb}. Since all of our calculations have assumed a low
 charge density, which requires $y=e^{-E_c/\kbt}\ll1$, which in turn
 implies $E_c/\kbt\gg1$, we can use the asymptotic
 expression\cite{Lamb} for the Lambert W function $W_0$ for large
 argument to simplify \rf{Tkt sol} to
 \beq k_B T_{KT}\approx{\sigma
   E_c\over2\ln\left({\sigma E_c\over K}\right)} \,.
 \label{Tkt sol approx}
 \eeq
 From this solution, it is clear that at least one bit of intuition
 about the effect of increasing the range of the potential (i.e.,
 increasing $\sigma$) is correct: as we increase $\sigma$ for fixed
 core energy $E_c$ and interaction strength $K$, the
 Kosterlitz-Thouless transition temperature $T_{KT}$
 grows. Specifically, it grows roughly linearly with $\sigma$, up to
 the extremely weak logarithmic dependence on $\sigma$ in the
 denominator of \rf{Tkt sol approx}. However, the most important point
 is that it remains finite, and that {\it all} super-Coulombic
 potentials in 2d undergo an unbinding transition in the same
 Kosterlitz-Thouless universality class as Coulombic (i.e.,
 logarithmic) potentials.

 Above this Kosterlitz-Thouless transition temperature, the charges
 are unbound, and, hence, the unconventional (power-law) Debye-Huckel
 screening we discussed in the previous subsection for the unconfined
 super-Coulombic case will also take place in the unbound phase. Since
 here we are in $d=2$, the result \rf{Usubfin} becomes
\beqn
U_{\text{eff}}(\br)&\approx&\frac{C^R_{DH}}{r^{4+\sigma}} \,.
\label{UKT}
\eeqn
The interaction strength $C^R_{DH}$ is no longer given by \rf{cdhdefR},
however, since all parameters will be renormalized by critical
fluctuations near the the Kosterlitz-Thouless transition. We defer a
calculation of the critical behavior of this constant to a future
publication.

\subsection{Unbinding transition for $d<2$}

For $d<2$, the Coulomb potential to which super-Coulombic bare
potentials ($\sigma > 0$) dielectrically screen is binding (that is,
it still grows with distance without bound like $r^{2-d}$).
Furthermore, {\it sub-Coulombic} potentials ($\sigma < 0$), are
dielectrically unscreened, and, when $\omega>0$ (which can only occur
in $d=1$), are also binding. For $\sigma >0$ (super-Coulombic)
potentials in $d=1$, the large distance behavior of the plasma (that
is, its behavior on length scales $r>\xi$) is that of a plasma with
linear ($U(r)\propto r$) interactions, while for $\sigma < 0$
(sub-Coulombic) potentials, $U(r)\propto r^\omega$ with
$0<\omega<1$. Such $d=1$ systems have been considered by
Kosterlitz\cite{Kost}, who showed that, while charges are bound at low
temperatures, there is an unbinding transition at a temperature $T_c$,
above which the system is screened. We can reproduce this results
using our sine-Gordon approach.

This transition in $d = 1$ is of a fairly conventional type. In particular, it
has a finite correlation length exponent. That is, the correlation
length $\xi_\text{deconfine}$ (which above $T_c$ is the screening length) diverges
according to
\beq \xi_\text{deconfine} \propto |T-T_c|^{-\nu} \,,
\label{xiunbind}
\eeq
in contrast to the exponential divergence of the correlation length at
the Kosterlitz-Thouless transition\cite{KT}. For small $\omega$, the
exponent $\nu$ is given by\cite{Kost} \beq \nu={1\over\sqrt{2\omega}}
\,.
\label{nuunbind}
\eeq
For super-Coulombic ($\sigma >0$) potentials, the interaction screens
down to Coulombic, which in $d=1$ is characterized by an effective
$\omega = 1$. Although for such large $\omega$ we lose quantitative
control of $\nu$, we expect that the power-law divergence of
$\xi_\text{deconfine}$ in \rf{xiunbind} remains valid, with a
$\nu(\omega)$, which is universal in the sense of depending only on
$\nu$.

As with the Kosterlitz-Thouless transition of the previous subsection,
here too, above the unbinding transition $T_c$ the charges are in a
plasma phase that will exhibit the unconventional power-law
Debye-Huckel screening, \rfs{Usubfin}, we derived in previous
subsection for the unconfined super-Coulombic case.

\section{Long-ranged XY models}
\label{secXY}
As noted in the Introduction, a significant motivation for our work
was originally driven by recent experimental realizations of spin
systems with power-law exchange, such as dipolar ultra-cold atomic,
ionic and molecular gases.  One may then expect that screening of a
super-Coulomb vortex gas will correspondingly lead to a modification
of the power-law exchange of the 2D XY model.  In this section we
critically examine this expectation and demonstrate that it is {\em
  not} realized.

\subsection{Failure of mapping to the super-Coulombic vortex gas}
Based on the mapping of the 2D XY model onto a Coulomb gas of
vortices\cite{KT}, one might be tempted to naively conclude that our
above screening results could be applied to a 2D XY model with
long-range exchange interactions -- that is, to a model in which unit
length two-component "spins" live on a two-dimensional lattice whose
sites are labelled by $\rv$.  The classical Hamiltonian for such a
system is:
\begin{eqnarray}
  H   &=& -\sum_{\rv,\rv'}\;
          J_{\rv,\rv'}\bS_\rv\cdot\bS_{\rv'}
      = -\sum_{\rv,\rv'}\;  J_{\rv,\rv'}\cos[\phi_\rv -
          \phi_{\rv'}]\;,\;\;\;\;\;\;\;\;\;
          \label{latticeH}
\end{eqnarray}
with the exchange coupling falling off algebraically with distance:
\beq
J_{\rv,\rv'}=J_0\left({a\over|\rv-\rv'|}\right)^{4-\sigma} \,,
\label{jdef}
\eeq
where the length $a$ is a lattice constant. Our choice of the
definition of $\sigma$ in the power-law exponent
$\alpha \equiv 4 -\sigma$ ($= 2 + d-\sigma$ in d-dimensions, that
differs from others in the literature e.g.,
$\alpha = d + \sigma_F$\;\cite{Fisher,translation}; see Appendix
\ref{AppendixGL}) is motivated by our definition of $\sigma$ in
\rf{Udef} and \rf{UFT}, as will become clear below.\cite{commentMap}

As famously shown by Kosterlitz and Thouless\cite{KT}, the {\it
  short-ranged} XY model can be mapped onto a Coulomb gas of vortices
by Taylor-expanding the cosine in \rf{latticeH} to quadratic order,
going to the continuum, and by including vortex topological defects by
writing
\beq
  \phi(\rv) =\sum_\alpha n_\alpha\phi^{\text{vortex}}(\rv -
  \rv_\alpha) \,,
 \label{many vortex field}
 \eeq
where we have defined
\beq
\phi^{\text{vortex}}(\rv)=\arctan(y/x) \,.
\label{single vortex field}
\eeq

Naively applying the same reasoning to the {\em long-ranged} XY model,
\rf{latticeH} one would (erroneously, as we will see below) conclude
that the resulting vortex gas would map onto the super-Coulombic
plasma \rf{Hlr} which we analyzed above, with the value of $\sigma$ in
\rf{Hlr} given by the $\sigma$ of (\ref{jdef}).  If this were correct,
then the universal screening result \rf{ussolvasym} would predict that
at scales longer than the crossover-to-Coulomb interaction length,
$\xi$, such a long-ranged XY model would reduce to the conventional,
short-ranged XY model, and in 2D would be expected to undergo a
Kosterlitz-Thouless phase transition.\cite{KT}

However, in fact, we know from the seminal work of Fisher
et. al.\cite{Fisher} that the transition in a long-ranged 2D XY model
is clearly {\it not} of the Kosterlitz -- Thouless type.  That paper
(which we review in Appendix \ref{AppendixGL}) studied the
complementary long-ranged Ginzburg-Landau $|\psi|^4$ model, in which
the angle $\phi$ in \rf{latticeH} is the complex phase of the ``soft
spin'' $\psi(\rv)\sim e^{i\phi}$. This model faithfully includes
vortex and spin-wave degrees of freedom.

The most dramatic difference between the KT and the transition found
in Ref.\onlinecite{Fisher} is that the low-temperature phase has
long-ranged ferromagnetic order, while the low-temperature
Kosterlitz-Thouless phase has only quasi-long-ranged order. Almost as
dramatic a difference is Fisher et. al.'s result that the
ferromagnetic correlation length in high temperature phase diverges
algebraically as the transition is approached from above, while it
diverges exponentially at the KT transition\cite{KT}. Thus, Fisher et.
al.'s\cite{Fisher} results (which can be tested for $\sigma=1$ in the
experimental system of Chen et. al.,\cite{Browaeys2022}) therefore
indicate a failure of the naive mapping of the long-ranged XY model
onto a super-Coulombic vortex plasma (which we have shown above {\it
  does} have a KT transition in 2D).

It is natural to ask why the mapping of the long-ranged XY model to a
gas of super-Coulombic vortices fails. The reason is the following: It
is clear that in a {\it long-ranged} XY model, the energy is dominated
by the interaction between spins that are far apart, for which, in the
presence of vortices, the difference $|\phi_\rv-\phi_{\rv'}|$ is
large, of order $2\pi$.  This therefore precludes the quadratic Taylor
approximation of the cosine in \rf{latticeH} and therefore thwarts the
naive mapping onto a quadratic Hamiltonian in $\phi_\rv$ and by
extension onto the super-Coulombic pairwise interacting model
\rf{Hlr}.

\subsection{Vortex-free harmonic model}
The lack of screening in the long-ranged XY model suggests that in the
long-range ordered phase\cite{Fisher}, vortices (expected to be
tightly confined in the ordered phase) may be neglected.  Therefore,
the spin-wave theory, obtained by expanding the model \rf{latticeH}
for small spin-waves $\phi_\rv-\phi_{\rv'}$ to quadratic order, with
the crucial implicit constraint of no vortices in $\phi_r$, is
expected to be accurately described by the continuum Hamiltonian,
\beqn
H&\approx&
\oh J_0 a^{-\sigma}\int_{\xv,\xv'}\;
\left({1\over|\rv-\rv'|}\right)^{4-\sigma}[\phi(\xv)-
\phi(\xv')]^2\,, \label{contH}\\
&\approx&\oh K a^{-\sigma}\int_\qv\;
q^{2-\sigma}|\phi_\qv|^2\equiv \oh\int_\qv\;
\hat{K}(\bq)q^2|\phi_\qv|^2\;.\;\;\;\;\;\;\;\;
\label{modelH}
\eeqn
%
 %
In the second line we defined $K\equiv C(\sigma) J_0$, 
with
$C(\sigma)={8\cos\left({\pi\sigma\over2}\right)\Gamma(\sigma)
  \Gamma^2\left({3-\sigma\over2}\right)\over2^\sigma(2-\sigma)(1-\sigma)\Gamma(3-\sigma)}$
an $O(1)$ numerical factor. In the last equality we have defined the
strongly wavevector-dependent ``spin-wave stiffness''
\beq \hat{K}(\bq)\equiv
\frac{2J_0}{a^{\sigma} q^2}\int_\rv
\frac{1-\cos(\qv\cdot\rv)}{r^{4-\sigma}}
\approx_{qa\ll 1}\frac{K}{(qa)^{\sigma}} \,,
\label{khat}
\eeq
which diverges as $\bq\to{\bf 0}$.  We note that for the
physically most relevant case of ``dipolar'' exchange,\cite{JunYe2013,
  SageRMP2019, Browaeys2022, Bakr2022} $\sigma = 1$ and the dispersion
in \rf{modelH} is $\propto|q|$, with $C(1)=2\pi$.

In this harmonic approximation, fluctuations of $\phi_\rv$ within the
ordered state are given by a simple Gaussian integral (or,
equivalently, the equipartition theorem) with the Hamiltonian
\rf{modelH},
\begin{eqnarray}
  \langle\phi^2(\rv)\rangle
  &=& \int \frac{d^2q}{(2\pi)^2}\frac{k_BT}{\hat{K}(\bq)q^{2}}
  = a^\sigma \int \frac{d^2q}{(2\pi)^2}\frac{k_BT}{K
       q^{2-\sigma}},\;\;\;\;\;\; \\
&=& \frac{k_BT}{2\pi\sigma K},
\label{phiRMS}
\end{eqnarray}
where we have used $1/a$ as the ultraviolet cutoff on the
$ \int d^2q$, and, for $\sigma > 0$ in the thermodynamic limit,
$L\rightarrow\infty$, neglected finite size corrections which are
smaller by a factor of $(a/L)^\sigma$.  Thus, for long-range exchange
  -- i.e., $\sigma > 0$ -- we find that $\phi_{\text{rms}}$ is finite
  and system-size $L$-independent in the thermodynamic limit.  This is
  clearly a reflection of the stabilizing effect of long-range
  interactions, with an effective exchange coupling
  $\hat K(q)\sim q^{-\sigma}$ that diverges at long length scales.  It
  is consistent with the 2D long-range order and nonzero magnetization
  $\langle\bS_\rv\rangle$ at low temperatures found by Fisher et.,
  al.\cite{Fisher} in their complementary ``soft spin'' description,
  and recently observed in two-dimensional Rydberg
  arrays.\cite{Browaeys2022}

The ordered phase is also characterized by a nontrivial longitudinal
susceptibility $\chi(h)$  for the response of the magnetization $M(h) = |\langle\bS_\rv \rangle|$
 to an applied field $h$. Perturbing the zero-field
Hamiltonian $H$ with $-\int_\rv{\bf h}\cdot\bS_\rv$ gives
$H_h = H - h\int_\rv\cos(\phi(\br))$\,.
In the ordered low-$T$ phase, this gives for the magnetization
\beqn
M(h)&=&|\langle\bS_\rv  \rangle|=\langle\cos(\phi(\br)) \rangle
\approx1-{\langle\phi^2(\rv) \rangle\over2}
 \nn\\
&=&1-\int \frac{d^2q}{(2\pi)^2} \frac{k_BT}{[K_0a^{-\sigma} q^{2-\sigma}+h]} \,.
\label{M}
\eeqn
The differential susceptibility $\chi(h)$ at small $h$
can now be determined by differentiating (\ref{M}):
\beqn
\chi(h)&\equiv&\left({\pp M\over\pp h}\right)_T
=\int \frac{d^2q}{(2\pi)^2} \frac{k_BT}{[K_0a^{-\sigma} q^{2-\sigma}+h]^2}\;,\nn\\
&\approx&
\left\{\begin{array}{ll}
         h^{-\mu}\;,\;\; \mbox{for $0 < \sigma <1$},\\
         \frac{k_BT a^2}{2\pi K_0^2}\ln\left({K_0\over a^2 h}\right) \;,\;\; \mbox{for $\sigma  = 1$},\\
      \chi_0\;, \;\;\;\;\;\mbox{for $1 < \sigma <2$},
        \end{array}\right.
\label{chih1}
\eeqn
where we have defined the exponent
\beq
\mu\equiv2\left(\frac{1-\sigma}{2-\sigma}\right)\;,\;\; \mbox{for $0 <
  \sigma <1$};,
\label{mu}
\eeq
and $\chi_0$ is a finite constant.

The qualitative distinction between the cases of $0 < \sigma\leq 1$
and $1 < \sigma <2$ comes from the infrared divergence and convergence
of the integral for $h=0$ in \rf{chih1}, respectively, which
correspondingly result in a divergent and finite susceptibility for
$h\rightarrow 0$. The former case for $0 < \sigma < 1$ corresponds to
a {\em sublinear} longitudinal magnetization response,
  \beq
  M(h)-M(0)\propto h^{\sigma/(2-\sigma)}.
  \eeq
For the case $\sigma=1$, as in the experiments of Bakr et. al.\cite{Bakr2022}, we get a logarithmic correction to the usual linear susceptibility, which is explicitly displayed in the middle line after the brackets in equation \rf{chih1}, while for $\sigma>1$, the linear susceptibility is finite.
   

\section{Results and conclusion}
\label{secConclude}
Motivated by the increasing number of experiments on long-ranged
interacting systems, we have studied two-component power-law
interacting super-Coulombic gases.  We derived in complementary ways --
via coarse-graining a sine-Gordon model and dielectric medium analyses
-- the screened inter-charge potential for such systems. Our key
striking finding is that dipole screening universally leads to the
standard $1/q^2$ Coulomb interaction, independent of the bare
potential, as long as it is longer ranged than Coulombic.  Along with
a number of regimes summarized in Fig. \ref{regimesFig}, we also
showed that, in contrast to naive expectations based on its
short-ranged counterpart, the 2D long-ranged XY model is, in fact,
{\em not} related to the super-Coulombic gas of vortices. It therefore
is {\em not} screened down to a short-ranged XY model. Instead, it
exhibits long-range ferromagnetic order with power-law bound vortices
at low $T$, and does {\em not} undergo a Kosterlitz-Thouless
transition.

We close by noting that the dielectric screening analysis is generic
and dimension independent.  It is intriguing to speculate that this
screening mechanism may explain the ubiquity of Coulombic potentials
in nature.\\

\noindent
{\em Note Added:} 
After this work was completed (with the key result obtained over 34
years ago!)  and in the long and delayed process of being written up,
we learned of a recent interesting paper, arXiv:2209.11810, Journal of
High Energy Physics, 2023(2), 1-25 (2022), {\em Villain model with
  long-range couplings} by Guido Giachetti, Nicolo Defenu, Stefano
Ruffo, and Andrea Trombettoni.\cite{Giachetti2022} There is overlap of
our results with those found in this nice work, although ours has a
somewhat broader scope.

In a separate development, we recently learned of an interesting paper
by Igor Herbut and Babak Seradjeh\cite{HerbutQED3} who study the
question we address, but in a very specific setting of a magnetic
monopole gas in QED-3, and argue that screening down to Coulomb
interaction ensures instability of the state to deconfinement of
monopoles and concomitant confinement of fermions. 

\emph{Acknowledgments}.  We thank Waseem Bakr for discussions and the
motivating experimental references. LR acknowledges support by the
Simons Investigator Award from The James Simons Foundation. LR also
thanks The Kavli Institute for Theoretical Physics for hospitality and
its participants for discussions (especially Nicolo Defenu and
co-authors of reference \cite{Giachetti2022}) during the workshop
"Exploring Non-equilibrium Long-range Quantum Matter", supported by
the National Science Foundation under Grant No. NSF PHY-1748958 and
PHY-2309135. JT thanks the IBM Research Almaden Lab for their
hospitality during a summer 1990 visit during which this work was
initiated, and the Max-Planck-Institut fur Physik Komplexer
Systeme(MPIPKS), Dresden, Germany, for their support and hospitality
during a visit at which a portion of the writing of this paper was
done.


\appendix
\section{Coarse-graining of a ``soft'' sine-Gordon model}
 
 Here we present the details of the renormalization group (RG)
 analysis of the``soft'' sine-Gordon model \rf{sgHlr}, to prove our
 claim that the super-Coulombic 2D vortex gas is universally screened
 to a Coulombic interaction.

 We begin by writing the sine-Gordon Hamiltonian \rf{sgHlr} of the
 main text as
 \beq H_{SG}=H_0+H_g \,,
 \label{Hsg}
 \eeq 
 where
 \begin{eqnarray}
 H_0=\ \oh \tilde
 K\int_{\qv}\; q^{2+\sigma}|\theta_{\qv}|^2 \,
\label{H0def}
\end{eqnarray}
is the quadratic part of the Hamiltonian, while
 \begin{eqnarray}
H_g= -g \int_\rv\cos(\theta_\rv ) \,
\label{Hgdef}
\end{eqnarray}
is the non-trivial part. Note that in \rf{Hgdef} we have kept only the
lowest harmonic in the Fourier series in \rf{sgHlr}, and have defined
$g\equiv g_1$
 
The RG\cite{JKKN} starts by separating the field
$\theta_\br=\theta^>_\br+\theta^<_\br$ into ``fast'' and ``slow''
components $\theta^>_\br$ and $\theta^<_\br$, where the ``fast''
component $\theta^>_\br$ only has support in the ``shell'' of Fourier
space $b^{-1} \Lambda \le |\bq| \le \Lambda$, where $\Lambda$ is an
``ultra-violet cutoff", while the ``slow'' component $\theta^<_\br$
has support in the ``core'' $0 \le |\bq| \le b^{-1} \Lambda$.  Here
$\Lambda$ is of order the inverse lattice constant $a^{-1}$ in a
lattice model, and for continuum models comparable to the microscopic
length scale on which our ``charges'' begin to show structure. Here
$b$ is an arbitrary rescaling factor. Later, to derive {\it
  differential} recursion relations, we will take $b = e^{d\ell}$ to
be close to $1$, with $d\ell \ll 1$.
 
We then derive an ``intermediate'' Hamiltonian $H_I(\theta^<)$ for the
slow degrees of freedom $\theta^<$ by integrating the Boltzmann weight
$Z^{-1}e^{-\beta H[\theta^<_\br,\theta^>_\br]}$ over the ``fast''
degrees of freedom $\theta^>_\br$, with $\beta\equiv1/\kbt$.  That is,
we write
 \bew
 \beq
Z_I^{-1} e^{-\beta H_I[\theta^<_\br]}=Z^{-1}\int
\prod_{\bq} d\theta^>_\bq\; e^{-\beta H[\theta^<_\br,\theta^>_\br]}
= Z_0 Z^{-1}e^{-\beta H_0[\theta^<_\br]}
\big\langle e^{-\beta
  H_g[\theta^<_\br,\theta^>_\br]}\big\rangle_0^>
\,,
\eeq
\ew
where $\int \prod_{\bq} d\theta^>_\bq$ denotes an integral over {\it
  only} the fast degrees of freedom $\theta^>_\bq$, and the symbol
$\langle\ldots \rangle_0^>$ denotes an average of $\ldots$ over those
fast modes using {\it only} the Boltzmann weight
$ Z_0^{-1} e^{-\beta H_0[\theta^>_\br]}$ for the purely quadratic part
of the Hamiltonian.  Since this Hamiltonian is quadratic, these
averages are straightforward to evaluate, being averages over a purely
Gaussian distribution.
 
As we will see below, this quadratic part of the Hamiltonian will be
drastically modified by the generation of the
$\oh\kappa \int_\rv(\grad\theta)^2$ term, which will eventually, after
a sufficient amount of renormalization group ``time'', come to
dominate over the $\tK$ term given above. In the discussion that
follows, we will initially ignore the effect of this generated term on
the quadratic Hamiltonian, and then we will discuss when this
assumption ceases to be valid.  The point at which that fails gives us
the crossover length $\xi$ between the super-Coulombic and Coulombic
interactions.
 
The next step of the RG is (purely for convenience) to rescale lengths
so as to restore the ultraviolet cutoff to its original value.  The
rescaling of wavevectors that accomplishes this is, obviously, the
change of variable
\beq \bq=b^{-1}\bq' \,.
\label{qrescaleRG}
\eeq
Due to the inverse relation between wavevectors $\bq$ and real space
coordinates $\br$, this implies the opposite scaling of coordinates:
\beq
\br=b\br' \,.
\label{rrescaleRG}
\eeq We choose {\it not} to rescale the real-space field $\theta_\br$,
in order to keep the coefficient of $\theta_\br$ in the the argument
of $\cos(\theta_\br)$ equal to $1$, i.e., to keep the periodicity
fixed at $2\pi$, corresponding to discrete integer charges of the
super-Coulomb gas.  The renormalization group now proceeds by
repetition of the above three RG steps (separating into fast and slow
fields, averaging over the fast fields, and rescaling).

We perform the averaging over $\theta^>_\br$ perturbatively in
$H_g= -g\int_\rv\cos\theta_\br$.  A standard second-order perturbation
analysis gives for the change $  \delta H $ in the effective Hamiltonian  for the remaining slow fields:
\begin{eqnarray}
  \delta H = \langle H_g\rangle_{0>} - \frac{\beta}{2} \langle
  H_g^2\rangle_{0>}^c + \cdots\;,
  \label{deltaH}
\end{eqnarray}
 The rescaling step of the RG gives,
\begin{equation}
  \tK(b) = \tK b^{-\omega} \,,\;\;\;\text{with}\;\; \omega = 2 - d + \sigma\;. 
\label{tKb}
\end{equation}
This relation is {\it exact}, since $\tK$ gets no corrections from the
perturbative coarse-graining, due to the non-analytic, $q^{2+\sigma}$
form of the $\tK$ term in \rf{H0def}.

The leading-order coarse-graining correction to the coupling $g$ comes
from the first term in \rf{deltaH}, which gives 
\begin{eqnarray}
\delta H=  \langle H_g\rangle^>_{0}
  &=& -g \int_\rv\langle\cos(\theta^<_\rv +
      \theta^>_\rv)\rangle^>_{0}\;\nn\\
   &=& -g  e^{-\oh \langle(\theta^>_\rv)^2\rangle^>_{0}}\int_\rv\cos\theta^<_\rv \,,
\label{deltaH1}
\end{eqnarray}
where the second equality follows from writing the cosine in terms of
complex exponentials, and then using the fact that,  for a zero-mean,
Gaussian field $\theta_>$ satisfies,
\beq
\langle e^{i \theta}\rangle=e^{-\oh\langle \theta^2\rangle} \,.
\label{exp ave}
\eeq
This then gives the transformation of $g$ under the RG,
\begin{eqnarray}
g_R & = &g b^d e^{-\oh G^>({\bf 0})} \nn\\
 & \approx&g b^d \exp\big(- S_d(\ln b)\kbt/[2(2\pi)^d\tK \Lambda^\omega]\big) \,,
 \\\nn
\label{g1b}
\end{eqnarray}
where the $b^d$ factor comes from the length rescaling in
$\int\,d^dr$. In \rf{g1b}, we have defined the momentum-shell
correlator
\begin{eqnarray}
  G^>({\bf r}-{\bf r'}) &=& \langle \theta_>({\bf r})\theta_>({\bf
                          r}')\rangle^>_0
  =\frac{k_BT}{\tK}\int^>_{\bf q}\frac{e^{i{\bf q}\cdot({\bf r}-{\bf r'})}}{ q^{2+\sigma}}\,,
  \nn\\
\label{G0}
\end{eqnarray}
with, for small $\ln b$,
$G^>({\bf 0}) \approx S_d(\ln b)\kbt/[(2\pi)^d\tK \Lambda^\omega]$,
with $S_d$ the surface hyper-area of a $d$-dimensional unit ball.

The crucial $\oh\kappa (\grad\theta)^2$ contribution arises from the
second order term $- \frac{\beta}{2} \langle H_g^2\rangle_{0>}^c$ in
\rf{deltaH}. This term is \bew
\begin{eqnarray}
\langle H_g^2\rangle_{0>}^c &=&g^2 
\int_{\rv,\rv'}\bigg(\langle\cos[\theta_<(\br)+\theta_>(\br)]\cos[\theta_<(\br')+\theta_>(\br')]\rangle-\langle\cos[\theta_<(\br)+\theta_>(\br)]\rangle\langle\cos[\theta_<(\br')+\theta_>(\br')]\rangle\bigg)
\,.\nonumber\\
\label{H2g}
\end{eqnarray}
\ew
Again writing the cosines in this expression in terms of complex
exponentials, performing the averages using the handy relation \rf{exp
  ave}, and gathering terms, we get \bew
\begin{eqnarray}
\langle H_g^2\rangle_{0>}^c &=&{g^2\over2}  
\int_{\rv,\rv'}\bigg\{\cos\bigg[\theta_<(\br)-\theta_<(\br')\bigg]\bigg(\exp\bigg[-{1\over2}\bigg\langle\bigg(\theta_>(\br)-\theta_>(\br')\bigg)^2\bigg\rangle\bigg]-\exp\bigg[-\bigg\langle\theta^2_>(\br)\bigg\rangle\bigg]\bigg)\nn\\
&+&\cos\bigg[\theta_<(\br)+\theta_<(\br')\bigg]\bigg(\exp\bigg[-{1\over2}\bigg\langle\bigg(\theta_>(\br)+\theta_>(\br')\bigg)^2\bigg\rangle\bigg]-\exp\bigg[-\bigg\langle\theta^2_>(\br)\bigg\rangle\bigg]\bigg)\bigg\}
\,.\nonumber\\
\label{H2g2}
\end{eqnarray}
\ew
Now using the facts that 
\beqn
\bigg\langle\bigg(\theta_>(\br)\pm\theta_>(\br')\bigg)^2\bigg\rangle&=&2\bigg\{\bigg\langle\bigg(\theta_>^2(\br)\bigg\rangle\pm\bigg\langle\theta_>(\br)\theta_>(\br')\bigg\rangle\bigg\}\nn\\
&=&2[G^>({\bf 0})\pm G^>({\bf r}-{\bf r'})] \,,
\label{dif ave}
\eeqn
we can rewrite \rf{H2g2} as
\begin{eqnarray}
\langle H_g^2\rangle_{0>}^c & \approx & \frac{1}{2}g^2 e^{-G^>({\bf 0})}
\int_{\rv,\rv'}\bigg[{\cal C}_+(|\rv-\rv'|)
\cos\big[\theta_<({\bf r})-\theta_<({\bf r'})\big]\nn\\&+&{\cal C}_-(|\rv-\rv'|)
\cos\big[\theta_<({\bf r})+\theta_<({\bf r'})\big]
\bigg]
\;,\nonumber\\
\label{H2g1}
\end{eqnarray}
where we have 
defined the kernels
\begin{equation}
  {\cal C}_\pm({\bf r})=e^{\pm G^>({\bf r})}-1  \,.
  \label{kernel}
\end{equation}

Because these kernals ${\cal
  C}_\pm(\rv)$ only have support from high momenta (specifically,
momenta near the UV cutoff $\Lambda \sim
1/a$), they are both short-ranged. Hence, we can Taylor-expand both
cosines in \rf{H2g1} in
$\br-\br'$. For the second term, this simply generates, to leading
order, a
$\cos[2\theta(\br)]$ term, which is a higher, less relevant harmonic
that we will neglect. For the first term, this expansion gives $\delta
H = \oh\kappa_R
\int_\rv(\grad\theta)^2$, with
$\kappa_R=\kappa+\delta\kappa$, where
$\kappa$ is the value of
$\kappa$ before we performed this step of the RG. For the very first
step of the RG, $\kappa=0$, but on subsequent steps,
$\kappa$ will be non-zero, precisely because of its generation as
outlined above. That calculation gives
\begin{eqnarray}
  \delta\kappa & \approx&
                             \frac{1}{8}g^2 \beta e^{-G^>({\bf 0})}\int d^2r  \,r^2{\cal C}_+(r), \nn\\
                  & \approx&
                             \frac{1}{16}g^2\beta \int d^2r \, r^2 G_>^2(r)\;,\nonumber\\
                  & \approx&\frac{ c_2   g^2\kbt}{
                             16\tK^2}\ln b\;,
\label{kappa2g2}
\end{eqnarray}
to leading order in $\ln b$. Here we have defined $c_2 \equiv
(2+\sigma)^2/(2\pi\Lambda^{2(2+\sigma)})$.  Note that
$c_2$ remains constant under renormalization, since on each step of
the RG we rescale lengths to restore the ultraviolet cutoff
$\Lambda$ to its original value.

In deriving this result (\ref{kappa2g2}), we have neglected the
contribution lowest order in $G^>({\bf q})$, since it vanishes at
$q=0$. This is because, by definition, $G^>({\bf q})$ only has support
at high momenta near the UV cutoff $\Lambda$. We then evaluated the
integral $I_2\equiv\int d^2r\, r^2 G_>^2(r)$ as follows. First we
expressed it as
%
\begin{equation}
I_2\equiv\int d^2r\,  r^2 G_>^2(r)=\int d^2r\,  |\br G_>(r)|^2 \,.
\label{Gint0}
\end{equation}
%
Then Fourier transforming, we can rewrite \rf{Gint0} as
\bew
\beq
I_2=\int^>_{\bf q}{\rm FT}_\bq\big[\br G_>(r)\big]\cdot
{\rm FT}_{-\bq}\big[\br G_>(r)\big] \,,
\label{Gint}
\eeq
\ew
where ${\rm FT}_\bq\big[{\bf V}(\br)\big]$ denotes the Fourier
transform of any position dependent vector ${\bf V}(\br)$, evaluated
at wavevector $\bq$.  We can thus express it in the convenient form,
\bew
\beq
{\rm  FT}_\bq\big[\br G_>(r)\big]=\int d^2r\, \br
G_>(\br)e^{-i\bq\cdot\br}=i\nabla_\bq\int d^2r\,
G_>(\br)e^{-i\bq\cdot\br}=i\nabla_\bq G_>(\bq)\;,
\eeq
\ew
where the vector operator $\nabla_\bq$ denotes the gradient with
respect to the vector $\bq$.  Given our earlier result \rf{G0} for
$G_>(\bq)$, one can see that
\beq i\nabla_\bq
G_>(\bq)=-{i(2+\sigma)\bq\over q^{4+\sigma}}
\left({\kbt\over\tK}\right)\,.
\eeq
Inserting this result, and the analogous expression for
${\rm FT}_{-\bq}\big[\br G_>(r)\big]$, into (\ref{Gint}) gives
\bew
\beq I_2=(2+\sigma)^2 \left({\kbt\over\tK}\right)^2\int^>_{\bf
  q}\left({1\over
    q^{6+2\sigma}}\right)={(2+\sigma)^2\over2\pi\Lambda^{2(2+\sigma)}}\left({\kbt\over\tK}\right)^2\ln b\,.
\label{Gint3}
\eeq
\ew
in $d=2$, or, in general $d$,
\beq
I_d={(2+\sigma)^2S_d\over(2\pi)^d\Lambda^{2(3+\sigma)-d}}\left({\kbt\over\tK}\right)^2\ln b
\equiv c_d\left({\kbt\over\tK}\right)^2\ln b\,.
\label{Gint4}
\eeq

Combining the above $d$-dimensional coarse-graining analysis with the
length rescaling \rf{rrescaleRG} gives the full recursion relations for the renormalized couplings $\tK_R$, $g_R$, and $\kappa_R$ in terms of the original couplings $\tK$, $g$, and $\kappa$ (and, of course, the rescaling factor $b$):
\bew
\begin{equation}
  \tK_R = \tK b^{-\omega}\;,\;\;\;  g_R=b^dg \exp\bigg(- \bigg[{h_d\kbt\over2\tK}\bigg]\ln b\bigg)\;,\;\;\; \kappa_R=b^{d-2}\big(\kappa+\frac{ c_d   g^2\kbt}{
                             16\tK^2}\ln b\big)\,,
\label{rrsfull}
\end{equation}
\ew
where we have  defined
\beq
h_d\equiv{S_d\Lambda^{-\omega}\over(2\pi)^d}\;.
\label{hddef}
\eeq

As is standard practice in RG calculations, we will now rewrite these
recursion relations in differential form. To do this, we will choose
the rescaling factor $b$ to be very close to one. Specifically, we
will take $b=1+d\ell$, with $d\ell$ differential. We also imagine
iterating the renormalization group for $n$ steps, and define a
renormalization group ``time'' via $\ell\equiv nd\ell$. We now take the
limit $d\ell\to0$ and $n\to\infty$ such that the product $nd\ell=\ell$
remains finite.

Doing all of this, using the fact that, in the $d\ell\to0$  limit, $\ln b\to d\ell$,  considering the $n+1$'st RG step, which takes us from $\ell=nd\ell$ to $\ell+d\ell$,  and, finally, expanding in $d\ell$, we can rewrite \rf{rrsfull} as
\beqn
  \tK(\ell+d\ell) &=& \tK(\ell)(1-\omega d\ell)+O(d\ell^2) \,,
  \label{tKdiff1}\\ \nn\\ g(\ell+d\ell)&=&g(\ell) \bigg(1+\bigg[d- h_d\kbt/[2\tK)]\bigg]d\ell\bigg)\nn\\
  &+&O(d\ell^2) \,,
  \label{gdif1}\\\nn\\
   \kappa(\ell+d\ell)&=&\kappa(\ell)\bigg(1+\bigg[(d-2) +\frac{ c_d   g^2\kbt}{
                             16 \kappa\tK^2}\bigg]d\ell\bigg) \,.
                             \label{kappadif1}
                             \nn\\
\label{rrsdif1}
\eeqn
Now subtracting $ \tK(\ell)$, $g(\ell)$, and $\kappa(\ell)$ from both sides of  \rf{tKdiff1}, \rf{gdif1}, and \rf{kappadif1}, respectively, dividing the resultant equations by $d\ell$, and taking the limit $d\ell\to0$, we obtain the differential recursion relations:
\bea
{d\tK\over d\ell}&=&-\omega\tK\;,
\label{Ktilderr}\\
{dg\over d\ell}&=&\left(d-{h_d\kbt\over2\tK}\right)g\;, 
\label{grr}\\
{d\kappa\over d\ell}&=&(d-2)\kappa + {c_d\kbt g^2\over 16\tK^2}\;.
\label{kapparr}
\eea
It is useful to define a dimensionless measure of temperature (or, equivalently, of the inverse
of the stiffness $\tK$),
\beq
t(\ell)\equiv{h_d\kbt\over2\tK(\ell)}\;,
\label{tdef}
\eeq
in terms of which the flow equations reduce to
\bea
{dt\over d\ell}&=&\omega t\;,
\label{trr}\\
{dg\over d\ell}&=&\left(d-t\right)g\;,
\label{grr2}\\
{d\kappa\over d\ell}&=&(d-2)\kappa + \frac{f_d}{\kbt} t^2 g^2\;,
\label{kapparr2}
\eea
where we have defined
\beq
f_d\equiv{(2+\sigma)^2(2\pi)^d\over 4 S_d\Lambda^{d+2}}\;.
\label{fddef}
\eeq
These recursion relations are straightforwardly solved, giving
\bea
t(\ell)&=&t_0e^{\omega\ell}\;,
\label{tsol}\\
g(\ell)&=&g_0e^{\ell d}
\exp{\bigg[-{t_0\over\omega}\left(e^{\omega\ell}-1\right)}\bigg]\;,
\label{gsol}
\eea
and
\bew
\beq
\kappa(\ell)={f_dt_0^2g_0^2\over\kbt}e^{(d-2)\ell}\int_0^\ell
d\ell' e^{(2\omega+d+2)\ell'}\exp{\bigg[-{2t_0\over\omega}\left(e^{\omega\ell'}-1\right)}\bigg]\;,
\label{kappasol}
\eeq
\ew
where we have neglected the subdominant transient in
$\kappa(\ell)$. Here $t_0$ and $g_0$ are the bare values of $t$ and
$g$. For $t_0\ll1$ and for
$\ell\gg{1\over\omega}\ln{\left(\omega\over t_0\right)}$ the integral
in this expression has already converged, allowing us to extend the
upper limit of integration to $\ell\rightarrow\infty$. This gives
  \bea
\kappa(\ell)&\approx&{f_dt_0^2g_0^2\over\kbt}e^{(d-2)\ell}\int_0^\infty
d\ell' e^{(2\omega+d+2)\ell'}
\exp{\bigg[-{2t_0\over\omega}e^{\omega\ell'}}\bigg]\;,
\label{kappasol1}\nonumber\\
&&\\
&\approx&{f_d t_0 g_0^2\over2\kbt}e^{(d-2)\ell}
\left({\omega\over2t_0}\right)^{1+(d+2)/\omega}\Gamma(2+(d+2)/\omega)\;,
\nonumber\\
&\equiv&\alpha_d\frac{t_0^{-(d+2)/\omega}
  g_0^2}{\kbt\Lambda^{d+2}}e^{(d-2)\ell}\;,
\label{kappasol2}
\eea
where $\alpha_d =  (2+\sigma)^2(\frac{\omega}{2})^{\frac{d+2}{\omega}+1}(2\pi)^d
  \Gamma(\frac{d+2}{\omega}+2)/(4 S_d)$ is a dimensionless, $O(1)$ constant.

Our solution \rf{tsol} for $t(\ell)$, combined with our expression \rf{tdef} relating $t(\ell)$ to $\tK(\ell)$, determines $\tK(\ell)$:
\beq
\tK(\ell)=\tK_0e^{-\omega\ell}\;,
\label{tKsol}
\eeq
where $\tK_0={(\kbt)^2\over KC(\sigma)}={(\kbt)^2a^\omega\over K}\times O(1)$ is the "bare" (i.e., $\ell=0$) value of $\tK$.

As mentioned above, we have ignored the effect of the generated
$|\nabla\theta|^2$ coupling $\kappa$ on the averages over $\theta_>$
that we perform in each step of the RG. This approximation will
clearly break down at the RG time $\ell$ once the
$\kappa(\ell) q^2|\theta(\bq)|^2$ part of the energy starts to become
comparable to the $\tK(\ell) q^{2+\sigma}|\theta(\bq)|^2$ piece that
we have kept for the $\bq$'s of the $\theta(\bq)$'s over which we are
averaging on each RG step.  Since those $\bq$'s are those near the
Brillouin zone boundary $|\bq|=\Lambda$, this leads to the condition
\beq
\kappa(\ell^*)\Lambda^2=\tK(\ell^*)\Lambda^{2+\sigma}\;,
\label{corrlengthcond1}
\eeq
where we have defined $\ell^*$ as the value of $\ell$ at which the
$\kappa$ term starts to become important, and we recall that $\Lambda$
is kept fixed under the RG transformation.

Using our solutions \rf{tKsol} and \rf{kappasol2} for $\tK(\ell)$ and
$\kappa(\ell)$ in this condition \rf{corrlengthcond1} enables us to
solve for $e^{\ell^*}$:
\bea e^{\ell^*}
&=&\left({\kbt\tK_0t_0^{d+2\over\omega}\Lambda^{d+2+\sigma}
    \over \alpha_d g_0^2}\right)^{1/\sigma}\;,\label{xiSGsmA}\\
&\approx&\left[\left({K\over\kbt}\right)^{{d+2\over\omega}-1}\left({1\over
      y^2}\right)\right]^{1/\sigma}\times O(1) \,.
\label{xiSGsm}
\eea
In writing the second equality, we have used equations \rf{tKdef}
and \rf{g1est} of the main text, evaluated in the "bare" ($\ell=0$)
system, to relate $\tK_0$ and $g_0\equiv g_1(\ell=0)$, respectively to the
original parameters of the super-Coulombic gas model. We have also
used equation \rf{tdef} evaluated at $\ell=0$ for $t_0$. Finally, in
the last step, we have taken the ultraviolet cutoff $\Lambda\sim1/a$.

As usual in the RG, we can associate with this renormalization group time $\ell^*$ a  crossover length scale
\bea
\xi &=& a e^{\ell^*}\nn\\
&=&a\left[\left({K\over\kbt}\right)^{{d+2\over\omega}-1}\left({1\over  y^2}\right)\right]^{1/\sigma}\times O(1) \,,
\label{xiSGsmB}
\eea
which is the result quoted in the main text.

A standard RG analysis implies that the physical value
$\kappa_{\text{phys}}$ of $\kappa$ -- that is, the value that will be
observed in experiment -- can be obtained from the renormalized value
$\kappa(\ell^*)$ by undoing the effect of all the rescalings we did in
the RG. This can be seen to imply
\bea
\kappa_{\text{phys}}&=&e^{(2-d)\ell^*}\kappa(\ell^*)\nonumber\\&=&
{t_0^{-{d+2\over\omega}}g_0^2\over\kbt\Lambda^{d+2}}
= \kbt \left({\kbt\over
    K}\right)^{d+2\over\omega}y^2a^{2-d} \times O(1)\,,\nonumber\\
      &=&\kbt \left({\kbt\over
    K}\right)^{d+2\over\omega} a^{2-d}e^{-2E_c/\kbt} \times O(1)\,,\nonumber\\
&=& \frac{(\kbt)^2}{K}\left({a\over\xi}\right)^\sigma a^{2-d} \times O(1)\,.
\label{kappaphys}
\eea
In the main text we drop the subsript ``phys'' on
$\kappa_{\text{phys}}$ and refer to this stiffness simply as $\kappa$.

The most important point about this result is that the renormalized
value $\kappa_{\text{phys}}$ of the parameter $\kappa$ is not zero,
even though it {\it is} zero in the original bare model. By simply
reversing the steps of the duality manipulations that lead from our
original super-Coulombic gas model to the sine-Gordon theory, it is
easy to see that the existence of a non-zero $\kappa_{\text{phys}}$
implies that the effective renormalized interaction between test
charges becomes Coulombic if the test charges are separated by a
distance greater than the crossover length $\xi$.  This is our most
important conclusion: a longer-ranged than Coulombic interaction
between charges {\it always} ``screens down'' to a purely Coulombic
interaction at sufficiently long distances, in all spatial dimensions.

Quantitatively, the effective interaction on those longer length
scales, beyond $\xi$, is given by:
\begin{equation}
  H = \oh\sum_{\rv,\rv'} n_\rv U_{\text{eff}}(\rv - \rv') n_{\rv'}+E_c\sum_\rv n^2_\rv\;,
  \label{Hlrcoul} 
\end{equation}
with $U_{\text{eff}}(\rv)$ given by
\beqn
U_{\text{eff}}(\rv)=U_{\text{Coulomb}}(\rv)=
\left\{\begin{array}{ll}
         {(\kbt)^2\over S_d\kappa} \, r^{2-d}\;,\;\; \mbox{$d>2$},\\\\
     - {(\kbt)^2\over2\pi\kappa}\ln\left({r\over a}\right)\;, \;\;\;\;\;\mbox{$d=2$}\,.
        \end{array}\right.
\label{chih}
\eeqn
Concommitent with this screening, we predict that the 2D super-Coulombic
plasma undergoes a KT transition at a temperature $T_{KT}$ that satisfies\cite{JKKN} 
\beq
 {\kbt_{KT}\over2\pi\kappa(T_{KT})}=4 \,.
 \label{KT cond}
 \eeq
Using our expression \rf{kappaphys} for $\kappa(T)$, evaluated in $d=2$, in this condition \rf{KT cond} gives
\beq
\left({K\over\kbt_{KT}}\right)^{4/\sigma}e^{2E_c/\kbt}=O(1) \,.
 \label{KT cond2}
 \eeq
whose solution is
 \beq
 \kbt_{KT}={\sigma E_c\over2W_0\bigg({\sigma E_c\over2Ka^\sigma}\bigg)}
 \label{KT sol}
 \eeq
 where $W_0(x)$ is the $k=0$ branch of the Lambert W function.
 
\section{Dielectric screening  analysis}

Here we present the details of the calculation of the electric
susceptibility, $\chi$ used in the main text to compute the screening
length $\xi$.  To this end we simply need to calculate the mean dipole
moment $\langle \pv\rangle$ density in the presence of a local
"electric field" $\E=-\grad U(\br)$, which we take to be uniform over
the distance of the separation of the pair. (It is straightforward to
show that higher charges make a far smaller contribution to the
susceptibility at low temperatures, and so may be ignored.)

The mean dipole moment $\langle \pv\rangle$ density can be calculated
for an isolated $n_\alpha=\pm 1$ pair of unit charges using simple
Boltzmann statistics. This calculation is straightforward if we are in
a range of parameters such that two conditions are met:
\noindent 1) The mean size (i.e., distance between its contituent
$\pm 1$ charges) of an isolated dipole pair is large compared to the
lattice constant $a$, so that we can treat the dipole in the continuum
approximation. This requires that \beq \left({K\over
    \kbt}\right)a^\omega\ll 1 \,.
\label{K cond}
\eeq
\noindent 2) The local density of dipoles must be small enough that each dipole pair can be treated in isolation. This requires that the charge fugacity 
\beq
y\equiv e^{-E_c\over\kbt}\ll1 \,.
\label{y cond}
\eeq
Both conditions (\ref{K cond}) and (\ref{y cond}) {\it can} be
satisfied in the temperature range
\beq E_c\gg\kbt\gg Ka^\omega \,.
\label{full cond}
\eeq
as long as $E_c\gg K a^\omega$.

One physical example of such a situation would be a system in which
the total number of positive and negative charges are separately
conserved, and the density of each is small. In such a system, one
would treat the core energy $E_c$ as a type of a ``Lagrange
multiplier'', whose value would have to be chosen to keep the
densities of the charges small. This would mean that $E_c$ would
become proportional to $\kbt$, so as to keep the charge density fixed
as temperature is varied. The fixed ratio of $E_c$ to $\kbt$ would
also have to be large if the density of charges was small. Thus we
would have the first part of the condition (\ref{full cond}) (i.e.,
$E_c\gg\kbt$) automatically satisfied at all temperatures. The second
condition (i.e., $\kbt\gg Ka^\omega$) could then be always satisfied
by raising temperature sufficiently.

In general spatial dimension $d$ the $i$'th component of the mean
dipole moment $\langle \pv\rangle$ density can be obtained as
follows. First, let us put our system on a hypercubic lattice of
lattice constant $a$. Now, considering a hypercubic sub-volume
$V_s=L_s^d$ of the system whose lateral extent $L_s\gg a$, but assume
$L_s$ is still small enough that, in a typical configuration there are
no dipoles in the volume at all. Since the maximum possible dipole
density is $y^2/a^d$, which applies when the two elements of the
dipole are separated by one lattice constant, we can therefore achieve
this by requiring that
\beq a^d\ll V_s\ll a^d/y^2
\label{vscond} \,.
\eeq 

We can now think of this sub-volume as being a system in the Grand
Canonical Ensemble, with the remainder of the system acting as the
heat and particle bath. Hence, the probability $P(\br)$ of having a
single pair of oppositely charged $\pm1$ dipoles with separation $\br$
in the sub-volume is
\beq P(\br)=\left({V_s\over
    a^d}\right)\exp(-(-U_0(\br)+2E_c)/\kbt)/\mathcal{Q} \,,
\label{prob1}
\eeq
where $\mathcal{Q}$ is the Grand Canonical Partition function for the
sub-volume $V_s$, and the factor $\left({V_s\over a^d}\right)$ is
simply the number of sites in the volume on which the first charge of
the pair can be placed. (Once that first charge is positioned, the
second charge can only be a displacement {\it vector} $\br$ away, so
there is no further choice as to where to put that charge.)

The Grand Canonical partition function is, as always, a sum of the
Boltzmann weight over configurations with different numbers of charges
in the volume $V_s$, starting with the configuration with {\it no}
charges, for which the Boltzmann weight is $1$ (since the energy of a
charge free state is zero). Hence we have
\beq
\mathcal{Q}=1+O\bigg[y^2\left({V_s\over a^d}\right)\bigg]\approx 1 \,,
\label{Qest}
\eeq
where the second approximate equality follows from the fact that, by
construction $y^2\left({V_s\over a^d}\right)\ll 1$, by the second
strong inequality of \rf{vscond}.

Using \rf{Qest} in \rf{prob1} gives
\beq
P(\br)\approx\left({V_s\over a^d}\right)e^{-[-U_0(\br)+2E_c\big]/\kbt} \,.
\label{prob2}
\eeq
It is also clear that the second strong inequality of \rf{vscond}
implies that it is extremely unlikely that there will be {\it more}
than one dipole in volume $V_s$, or that there are dipoles made of
higher than unit charges.  It also implies that higher moment
configurations, such as, e.g., quadrupoles, which can be thought of as
configurations with more than one dipole pair close to each other, are
also highly unlikely, and therefore also negligible.  Hence, the only
configuration that makes an appreciable contribution to the total
dipole moment ${\bf \mathcal{P}}$ is the single dipole configuration
whose probability we just calculated. (The {\it zero} dipole state
obviously contributes nothing to the dipole moment.)

Since the dipole moment of a single pair is just the vector $\br$
separating the two charges, we have
\beqn 
\langle {\bf \mathcal{P}}\rangle&=&\sum_\br \br P(\br) \nn\\&\approx& y^2\left({V_s\over a^d}\right)
 \sum_\br\exp\left(-{(K(r/a)^\omega}-\E\cdot\rv)\over \kbt\right)\br\,, \nn\\
\label{dipole0}
\eeqn
where we have added an external "electric field" energy
$ \sum_\br\E\cdot\rv$. Our goal is to calculate the susceptibility to
this applied field.

Dividing both sides of \rf{dipole0} by the volume $V_s$ of our
sub-region gives the mean dipole moment per unit volume
$\langle {\bf p}\rangle$:
\beqn 
\langle {\bf \mathcal{{\bf p}}}\rangle
&=&V_s^{-1} \sum_\br \br P(\br)\nn\\
&\approx&
\left({y^2\over a^{2d}}\right) \int d^dr \exp\left(-{(K(r/a)^\omega}-\E\cdot\rv)\over \kbt\right)\br \,,\nn\\
\label{dipolevol}
\eeqn
where in the second equality we have used our expression \rf{prob2}
for $P(\br)$, and gone over to the continuum limit in the usual way,
i.e., via the replacement of sums over $\br$ with integrals over $\br$
using the substitution $\sum_\br\to a^{-d}\int d^dr$

The $i$'th component of this vectorial equation therefore reads
\beqn 
\langle p_i\rangle&=&\langle r_i\rangle=\left({y^2\over a^{2d}}\right)
\int\exp\left(-{(K(r/a)^\omega}-\E\cdot\rv)\over \kbt\right)r_i\,d^dr
\nn\\
&\approx&\left({y^2\over a^{2d}\kbt}\right)\int\exp\left(-{ K (r/a)^\omega\over \kbt}\right)r_ir_jE_j\,d^dr \,,
\label{dipole2}
\eeqn
where in the second equality we have expanded to linear order in $\E$.

Taking advantage of the rotational symmetry of the Boltzmann factor
$\exp\left(-{K(r/a)^\omega\over \kbt}\right)$, we can replace $r_ir_j$
in the integral by its angular average ${r^2\over d}\delta_{ij}$,
which reduces \rf{dipole2} to 
\beqn
\langle p_i\rangle&=&\bigg[{y^2\over d a^{2d}\kbt}\int\exp\left(-{K(r/a)^\omega\over \kbt}\right)r^2\,d^dr\bigg] E_i  \,.\nn\\
\label{dipole3}
\eeqn
That is, we find $\langle \bp\rangle=\chi \E$, 
with  the susceptibility $\chi$ given by
\beqn
\chi&=&{y^2\over da^{2d}\kbt}\int\exp\left(-{K(r/a)^\omega\over \kbt}\right)r^2\,d^dr
\nn\\
&=& {S_d\Gamma\left({d+2\over\omega}\right)y^2\over \omega d\kbt a^{d-2}}
\left({\kbt\over K}\right)^{\left({d+2\over\omega}\right)}
\nn\\
&=&{y^2\over \kbt a^{d-2}}\left({\kbt\over K}\right)^{\left({d+2\over\omega}\right)}\times O(1)\,.
\label{chi}
\eeqn
Using this result in our expression \rf{ussolvasym} for the crossover
length $\xi$ between super-Coulombic and Coulombic interactions gives
\beq
\xi=a^{\omega\over\sigma}\left(K\chi\right)^{-{1\over\sigma}}=a\left[\left({K\over\kbt}\right)^{\left({d+2\over\omega}\right)-1}\left({1\over
      y^2}\right)\right]^{1\over\sigma}\times O(1)
\label{xidielectric}
\eeq
in perfect agreement with the result (\ref{xiSGsmB}) obtained earlier
by the renormalization group analysis of the sine-Gordon theory.


\vspace{.2in}

 \section{Computational details of  Fourier
    transforms}
    
  \label{AppendixFT}
  In this appendix we demonstrate:

  \noindent 1) Equation \rf{UFT} of the main text, by showing that the
  Fourier transform of
  \beq U_0(q)=C(\sigma, d)K/q^{2+\sigma}
  \label{uq}
  \eeq
  back to real space does indeed give the bare interaction potential
  given by \rf{Udef} of the main text,
  \beq U_0(r) = - K (r/a)^\omega
  \sep \omega=2-d+\sigma \,,
  \label{ur}
  \eeq
  \vspace{.1in}
and 

\noindent 2) that the asymptotic long-distance tail of the real-space
Debye-Huckel screened effective potential is power-law
$U_{\text{eff}}(r) = C_{DH}(\sigma, d)/r^{d+2+\sigma}$. In the
process, we also calculate the constants $C_{DH}(\sigma, d)$ and
$C(\sigma, d)$. Our result for the former implies overscreening for
all super-Coulombic potentials, as discussed in the main text.
  
\subsection{Fourier transform of the bare potential when it is
  (naively) binding (i.e., $\omega>0$)}
  
Point 1) is slightly subtle to derive, because, in fact, the Fourier
transform of \rf{uq} back to real space contains, in addition to the
bare potential $U_0(r)$ given by equation \rf{ur}, an additive
constant which diverges for all binding potentials (that is, all
potentials with $\omega>0$) like $L^{\omega}$, where $L$ is the linear
spatial extent of the system in real space. Fortunately, this additive
constant, which also depends on the macroscopic shape of the system,
has no effect on the physics of the problem, other than (in the
grand-canonical ensemble) to enforce the constraint of overall charge
neutrality, which we expect to hold on physical grounds for confining
potentials. Once charge neutrality is enforced, this additive constant
drops out of the problem, as we will show below.
 
Since we need to consider a finite system in order to get a finite
answer, we begin by formulating Fourier transforms in a finite system.
We will take our system to be a $d$-dimensional rectilinear slab with
$d-1$ of its edge lengths given by $L_\perp$, and a single edge length
$L_z$, which need not equal $L_\perp$ in general. (This will enable us
to investigate the aspect ratio dependence of the aforementioned
additive constant.)

For computational convenience, we will use periodic boundary
conditions on this rectangular box, and define the Fourier transformed
charge density $n(\bq)$ via
\beq
n(\bq) = \frac{1}{\sqrt{V}}\int d^d r e^{-i \bq\cdot\br} n(\br)\;,
\label{ftdef}
\eeq
where 
\beq
V=L_\perp^{d-1}L_z
\label{V}
\eeq
is the (hyper) volume of the system, and the values of $\bq$
allowed by our periodic boundary conditions are
\beq
\bq = \frac{2\pi n_z}{L_z}\zh + \frac{2\pi}{L_\perp} \hat {\bf n}_\perp
  \,,
\label{qall}
\eeq
where all of the components of
\beq
\hat{\bf n} = n_z\zh + \hat{\bf n}_\perp
\label{ndef}
\eeq
are integers. We exclude $\bf{0}$ from the allowed values of $\bn$ to
exclude $\bq=\bf{0}$ from the sum over $\bq$.

We now take the Hamiltonian to be
\beq
H   = \oh \sum_\bq \frac{C(\sigma,d) K}{q^{2+\sigma}}|n(\qv)|^2
+ E_c\sum_\br n_\br^2\;,
\label{Hft}
\eeq
and see if we can make a choice of $C(\sigma,d)$ that recovers our
potential \rf{Udef}, \rf{ur}, plus an additive constant.  Inserting
\rf{ftdef} into \rf{Hft}, we find that to recover \rf{Udef}, we must
have
\beq
U_0(\br) = \frac{C(\sigma,d)K}{V}\sum_\bq \frac{e^{i \bq\cdot\br}}{q^{2+\sigma}}
\,.
\label{Uft}
\eeq

We now rewrite the sum in this expression as
\bea
S(\br) &=& \frac{1}{V}\sum_\bq \frac{e^{i 
    \bq\cdot\br}}{q^{2+\sigma}},\\
&=& \frac{1}{V}\sum_\bq \frac{1}{q^{2+\sigma}} +
\frac{1}{V}\sum_\bq \frac{(e^{i \bq\cdot\br}-1)}{q^{2+\sigma}}
\,.
\label{sum}
\eea
The first term in this expression (which we call $S_0$) is given
by
\bea
S_0 &=& \frac{1}{V}\sum_\bq \frac{1}{q^{2+\sigma}}\;,\\
&=& \frac{1}{L_\perp^{d-1}L_z}\sum_{\hat{\bf n}\neq 0}
\left[\frac{1}{n_z^2(2\pi/L_z)^2 +
    n_\perp^2(2\pi/L_\perp)^2}\right]^{1+\sigma/2}\;,\nonumber\\
&&\\
&=&\frac{L_\perp^\omega f^{1+\sigma}}{(2\pi)^{2+\sigma}}g(f)\,.
\label{s0def}
\eea
where we have defined aspect ratio
\beq
f={L_z\over L_\perp}
\label{fdef}
\eeq
and
\beq
g(f) = \sum_{\hat{\bf n}\neq 0}
\left[\frac{1}{n_z^2 + f^2  n_\perp^2}\right]^{1+\sigma/2}
\,.
\label{gdef}
\eeq
Note that the sum in \rf{gdef} converges as $\bn\to\infty$ if and only
if (iff) $\omega>0$, which is the case we are considering here.

Note also that the dependence of $S_0$ on $f$ demonstrates the shape
dependence of this additive constant. Such shape dependence also
occurs due to dipolar interactions in ferromagnets, as here, this is
due to the long-ranged nature of the interaction.

The two most important points to note about $S_0$ are: (i) it is
constant, and (ii) it diverges as system size $L_\perp\to\infty$
(keeping $f$ fixed).  Point (ii) might appear quite
alarming. Fortunately, point (i) renders the $S_0$ term almost
completely unimportant. Indeed, one can easily see that it leads to a
term in the total Hamiltonian
\bea
 H_{S_0} &=& \oh S_0\int_{\rv,\rv'} n(\rv) n(\rv')\;,\\
&=& \oh S_0\left(\int_\rv n(\rv)\right)^2 =  Q^2 S_0\;, 
\eea
where 
\beq
Q\equiv\int_\br \, n(\br)
\label{qdef}
\eeq
is the total charge in the system.  Because the charges are quantized,
any non-zero $Q^2$ must have a magnitude of at least $1$. Hence, any
non-neutral configuration of charges has a positive definite energy
cost that diverges as $L_\perp^\omega$, and, therefore, has a zero
Boltzmann weight. That is, the effect of this $S_0$ term is to enforce
total charge neutrality, which is a physical constraint we expect in a
system with interactions that diverge at large $\br$.

Once the constraint of charge neutrality, $Q=0$, is enforced, the
$S_0$ term drops out of the Hamiltonian entirely. Hence, we are left
with the model \rf{Hlr}, with $U_0(\rv)$ given by the second term in
\rf{sum}.

The sum in this term converges as the system size
$L_{\perp,z}\to\infty$ (at fixed $f = L_z/L_\perp$), as we will show
in a moment. This enables us to replace the sum on $\bq$ with an
integral over $\bq$ by the usual replacement
\beq
\frac{1}{V}\sum_\bq \ldots \rightarrow
\int\frac{d^d q}{(2\pi)^d}\ldots\;.
\label{sumtoint}
\eeq 
Doing this gives
\beq
U_0(\br) = C(\sigma,d)K\int\frac{d^d q}{(2\pi)^d}
\frac{e^{i \bq\cdot\br}-1}{q^{2+\sigma}}\,.
\label{U0.1}
\eeq
Choosing a (hyper)-spherical coordinate system for $\bq$ with its polar axis aligned along $\br$, and evaluating the integral over the $d-2$ angular co-ordinates other than the polar angle $\theta$ gives
\bea
U_0(\br) &=& C(\sigma,d) K\frac{S_{d-1}}{(2\pi)^d}\int_0^\pi
(\sin\theta)^{d-2} d\theta\nonumber\\
&\times&\int_\epsilon^\infty \frac{(e^{i q
    r\cos\theta}-1)}{q^{1+\omega}}dq\,,
\label{U0.2}
\eea
where $S_{d-1}$ is the surface area of a $d-1$-dimensional unit ball,
and $\epsilon=O(L^{-1})$ (for simplicity focussing on aspect ratio
$f = 1$) is an infra-red cutoff, crudely reflecting the discrete
nature of the $\bq$\Õs in our finite system. We will eventually
consider the thermodynamic limit $L\to\infty$, corresponding to
$\epsilon\to0$, and will show that this limit is well-defined and
finite for all $\omega<2$.

Integrating over $q$ in \rf{U0.2} by parts gives
\bea
&&\int_\epsilon^\infty \frac{(e^{i q
    r\cos\theta}-1)}{q^{1+\omega}}dq= -\omega^{-1}q^{-\omega}
  \left(e^{i q  r\cos\theta}-1\right)\left.\right|^\infty_\epsilon\nonumber\\
 &&+ i \omega^{-1} r \cos\theta \int_\epsilon^\infty \frac{e^{i q
    r\cos\theta}}{q^{\omega}}dq\,.
\label{byparts1}
\eea
The first term in this expression vanishes at infinity, while its
lower limit, for small $\epsilon$ (that is, large system size $L$) can
be evaluated by Taylor-expanding the complex exponential, giving
\bea
 &&q^{-\omega}\left(e^{i q
    r\cos\theta}-1\right)\left.\right|^\infty_\epsilon
=- i \epsilon^{1-\omega} r\cos\theta
+ \oh \epsilon^{2-\omega} r^2 \cos^2\theta
\,.\nonumber\\
&&
\label{eps1}
\eea
The first term in this expression vanishes when integrated over the
polar angle $\theta$.  The second term vanishes in the thermodynamic
limit $\epsilon\to 0$ for all $\omega<2$. Hence, for all $\omega<2$,
in the limit $\epsilon\to0$, the first, boundary term in \rf{byparts1}
can be dropped, leaving
\bea
U_0(\br) &=& i C(\sigma,d) K\frac{S_{d-1}}{\omega(2\pi)^d} r\int_0^\pi
\sin^{d-2}\theta\cos\theta d\theta\nonumber\\
&\times&\int_\epsilon^\infty \frac{e^{i q r\cos\theta}}{q^{\omega}}dq\,.
\label{U0.3}
\eea
This is as far as we can go for general $\omega$ in the range
$0<\omega<2$. To go further, we must separately deal with the cases
$0 <\omega < 1$ and $1 < \omega < 2$.

\subsubsection{$0 < \omega < 1$}
\label{0<omega<1}

\begin{figure}
    \centering
    \includegraphics[width=1\linewidth]{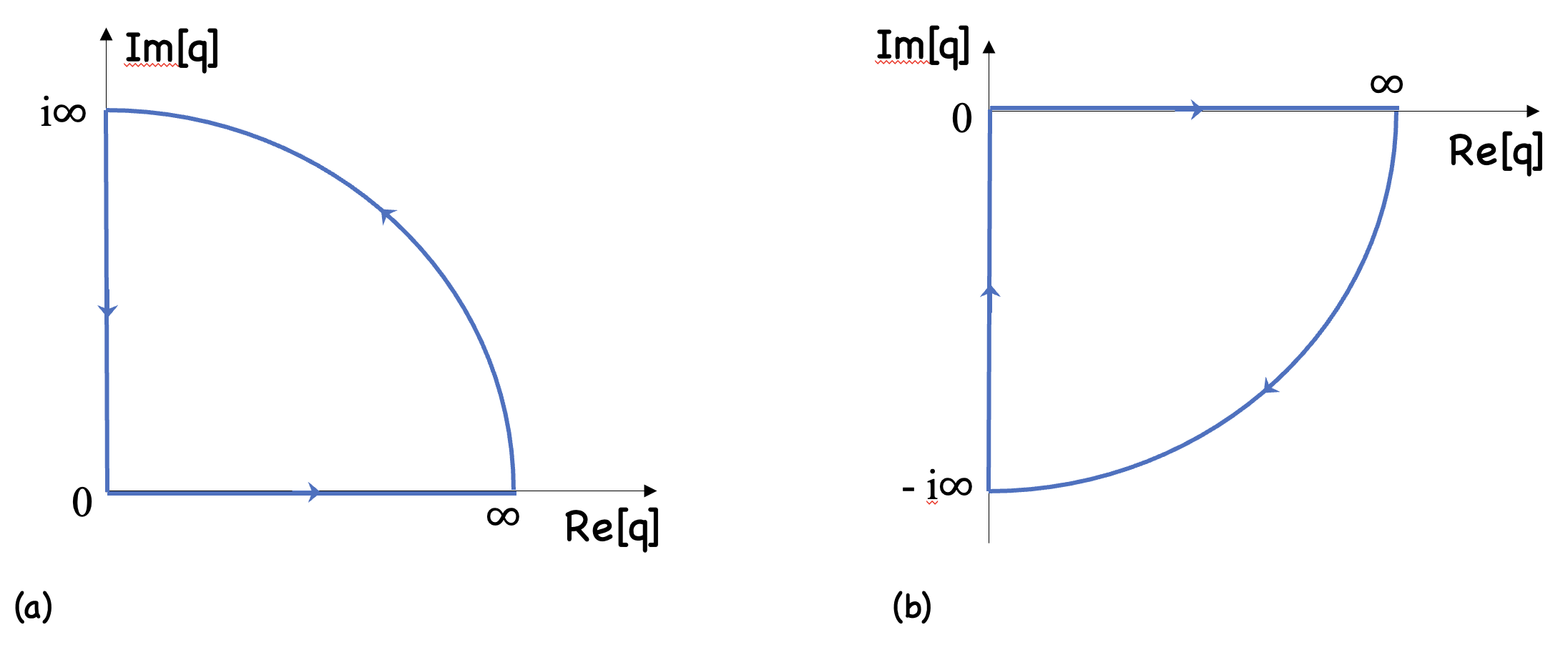}
    \caption{Integration contours in the complex $q$ space for
      $\cos\theta > 0$ in (a) and for $\cos\theta < 0$ in (b). }
    \label{contourIntegralFig}
  \end{figure}

For $\omega<1$, the integral over $q$ in \rf{U0.3} obviously
converges as $q\to0$.  Hence, we can take the limit
$\epsilon\to0$. This leaves us with the task of evaluating
\beq
I(r\cos\theta) =\int_0^\infty \frac{e^{i q r\cos\theta}}{q^{\omega}}dq\,.
\label{Idef<1}
\eeq
This can be done by rotating the contour in the complex plane, as
illustrated by Fig.\ref{contourIntegralFig}. This rotation can be done
since the closed contour does not enclose any poles and $q^{-\omega}$
vanishes as $q\to\infty$, making the integral over the quarter circle
in these contour integrals vanish. The vanishing of $q^{-\omega}$ as
$q\to\infty$ also ensures the convergence of the integral along the
real axis, since an extra power of convergence is gained from strongly
oscillatory factor.

 For $\cos\theta > 0$, (i.e., for $0<\theta<\pi/2$), we close the
 contour in the upper right quadrant \ref{contourIntegralFig}(a) and obtain
 \beq
I(r\cos\theta) =\int_0^{i\infty} \frac{e^{i q r\cos\theta}}{q^{\omega}}dq\,.
\label{w<1,cos t>0.1}
\eeq
Changing variable of integration from $q$ to $t$, defined by
$i t \equiv q r\cos\theta$, we obtain
\beq
I(r\cos\theta) = i e^{-i\omega\pi/2}(\cos\theta)^{\omega-1}
\Gamma(1-\omega) r^{\omega-1}\,.
\label{w<1,cos t>0.2}
\eeq
The calculation for $\cos\theta <0$, (i.e., for $\pi/2<\theta<\pi$) is
almost identical, except that we now close the contour in the lower
right quadrant, \ref{contourIntegralFig}(b) and obtain
\beq
I(r\cos\theta) = -i e^{ i\omega\pi/2}(-\cos\theta)^{\omega-1}
\Gamma(1-\omega) r^{\omega-1}\,.
\label{w<1,cos t<0.1}
\eeq
Using these results \rf{w<1,cos t>0.2} and \rf{w<1,cos t<0.1} in \rf{U0.3}, we obtain
\bea
U_0(\br) &=& -K r^\omega
\frac{C(\sigma,d)  S_{d-1}\Gamma(1-\omega)}{\omega (2\pi)^d}\nonumber\\
&&\times\left[-e^{-i\omega\pi/2}\int_0^{\pi/2}(\sin\theta)^{d-2}
 (\cos\theta)^{\omega}d\theta\right.\nonumber\\
&&\left.+  e^{i\omega\pi/2}\int_{\pi/2}^\pi(\sin\theta)^{d-2}
  (-\cos\theta)^{\omega -1} \cos\theta \,d\theta\right]
  \nn\\
  \;.
\label{w>1.5}
\eea
With the change of variables $\phi=\pi-\theta$ in the second integral
over $\theta$, we see that it is equal to minus the first integral,
which is tabulated\cite{GS}. We therefore choose $C(\sigma,d)$ such that
\bea
&&\frac{C(\sigma,d)\cos(\omega\pi/2)\Gamma(1-\omega) \Gamma((\omega+1)/2)
\Gamma((d-1)/2)S_{d-1}}{\omega (2\pi)^d\Gamma((2+\sigma)2)}\nonumber\\
&=&a^{-\omega}\;,
\label{Cw<1}
\eea
and thereby obtain equations \rf{Udef} and \rf{UFT} of the
Introduction, with
\beq
C(\sigma,d)=\frac{ a^{-\omega}\omega (2\pi)^d\Gamma((2+\sigma)2)}{\cos(\omega\pi/2)\Gamma(1-\omega) \Gamma((\omega+1)/2) 
\Gamma((d-1)/2)S_{d-1}} \,.
\label{csd0}
\eeq
Using 
\beq
S_{d-1}={2\pi^{\left({d-1\over2}\right)}\over\Gamma\left({d-1\over2}\right)}
\label{sd-1}
\eeq
in \rf{csd0}, we obtain
\beq
C(\sigma,d)=\frac{2^{d-1}\pi^{(d+1)/2}\omega \Gamma(2+\sigma)}{a^\omega\cos(\omega\pi/2)\Gamma(1-\omega) \Gamma((\omega+1)/2)} \,.
\label{csd}
\eeq

\subsubsection{$1< \omega < 2$}

For $\omega > 1$, the integral in \rf{U0.3} clearly does {\it not}
converge as $\epsilon\to0$. To overcome this problem, we first
integrate by parts, obtaining
\bea
&&\int_\epsilon^\infty\frac{e^{i q r\cos\theta}}{q^{\omega}}dq
  = \frac{1}{1-\omega}q^{1-\omega}
  e^{i q  r\cos\theta}\left.\right|^\infty_\epsilon\nonumber\\
  &&- \frac{i}{1-\omega} r \cos\theta
\int_\epsilon^\infty e^{i q  r\cos\theta}q^{1-\omega}dq\,.
\label{w>1.1}
\eea
The first term inside the square brackets in this expression now
vanishes at infinity, while at the lower limit, for very small
$\epsilon$, it becomes $\epsilon^{1-\omega}$. Since this is
independent of $\theta$, when multiplied by $\cos\theta$ and
integrated over $\theta$ it vanishes as in \rf{U0.3}. Thus we are
left with the second term, which now converges
as $\epsilon\to0$ for $\omega<2$. We therefore take the limit
$\epsilon\to0$ in this remaining term and obtain
\bea
U_0(\br) &=& C(\sigma,d) K\frac{S_{d-1}}{\omega(1-\omega)(2\pi)^d} r^2
\int_0^\pi\sin^{d-2}\theta\cos^2\theta d\theta\nonumber\\
&\times&\int_0^\infty e^{i q r\cos\theta}q^{1-\omega}dq\,.
\label{w>1.2}
\eea
The integral over $q$ in this expression can now be evaluated by exactly the same sort of complex contour techniques that we used in the previous subsection, because the integrand vanishes for $q\to\infty$ since $\omega>1$. We thereby obtain
\beq
\int_0^{i\infty} \frac{e^{i q r\cos\theta}}{q^{\omega-1}}dq
 = - e^{-i\omega\pi/2}(r\cos\theta)^{\omega-2}
\Gamma(2-\omega)\,,
\label{w>1.3}
\eeq
for $\cos\theta>0$, and
\beq
\int_0^{-i\infty} \frac{e^{i q r\cos\theta}}{q^{\omega-1}}dq
 = - e^{i\omega\pi/2}(-r\cos\theta)^{\omega-2}
\Gamma(2-\omega)\,,
\label{w>1.4}
\eeq
for $\cos\theta<0$.  Using these results in \rf{w>1.2} gives
\bea
U_0(\br) &=& K r^\omega
\frac{C(\sigma,d)  S_{d-1}\Gamma(2-\omega)}{\omega (1-\omega) (2\pi)^d}\nonumber\\
&&\times\left[e^{-i\omega\pi/2}\int_0^{\pi/2}(\sin\theta)^{d-2}
 (\cos\theta)^{\omega}d\theta\right.\nonumber\\
&&\left.+  e^{i\omega\pi/2}\int_{\pi/2}^\pi(\sin\theta)^{d-2}
  (-\cos\theta)^{\omega -2} \cos^2\theta \,d\theta\right]
  \nn\\
  \;.
\label{Uanglew<1}
\eea
With the change of variables $\phi=\pi-\theta$ in the second integral
over $\theta$, we see that it is equal to the first integral,
which is tabulated\cite{GS}. We thereby obtain exactly our earlier
result \rf{csd} for $C(\sigma,d)$,  after using the relation
$\Gamma(2-\omega)=(1-\omega)\Gamma(1-\omega)$, thereby extending the
validity of the result to  $0<\omega < 2$. 

\subsection{Non-binding potentials ($\omega<0$)}

For non-binding potentials ($\omega<0$),  it is possible to directly calculate the Fourier transform from real space to Fourier space:
\bea
U_0(\bq) &=& \int d^d r  U_0(\br)  e^{-i \bq\cdot\br}\;,\nonumber\\
&=& -\frac{K}{a^\omega}\int d^d r r^\omega e^{-i \bq\cdot\br}\;.
\label{FTdir1}
\eea
Once again, different values of $\omega$ require different approaches
to perform this integral.

\subsubsection{$-2< \omega < 0$}

In this case, the integral is most conveniently done in Cartesian
coordinates, with one axis, which we will call $r_\parallel$, running
along $-\bq$, and the other $d-1$ axes, which we will denote by
$\br_\perp$ running perpendicular to $\bq$. With this choice of
coordinates, the expression \rf{FTdir1} for $U_0(\bq)$ becomes
\beq
U_0(\bq) = -\frac{K}{a^\omega}\int d^{d-1} r_\perp
\int_{-\infty}^\infty d r_\parallel
(r_\parallel^2+r_\perp^2)^{\omega/2} e^{i q r_\parallel}\,.
\label{-2<w<0.1}
\eeq
The integrand in this expression considered as a function of complex
$r_\parallel$ has a branch cut along the positive imaginary axis,
running from $ir_\perp$ to $i\infty$.  We close the contour in the
complex plane around this branch cut, as illustrated in Fig.
\rf{contourIntegralFig2}.
\begin{figure}
    \centering
    \includegraphics[width=0.9\linewidth]{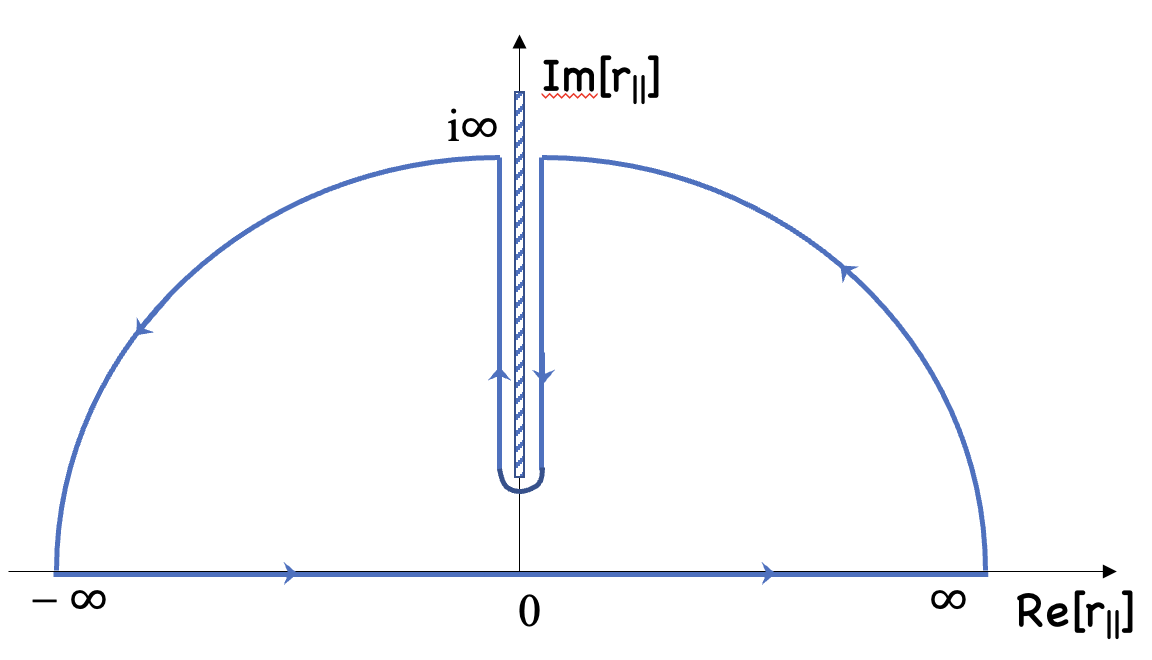}
   \caption{Integration ontour in the complex $r_\parallel$ plane. }
    \label{contourIntegralFig2}
  \end{figure}
  Because the integrand vanishes as $r_\parallel\to\infty$ for
  $\omega<0$, we can close this contour in the complex plane with
  impunity.  We thereby obtain
\bea
I(r_\perp) &=&\int_{-\infty}^\infty d r_\parallel
(r_\parallel^2+r_\perp^2)^{\omega/2} e^{i q r_\parallel}\,,\\
&=&\int_{i r_\perp}^\infty d r_\parallel 
(r_\parallel^2+r_\perp^2)_+^{\omega/2} e^{i q r_\parallel}\nonumber\\
&&-\int_{i r_\perp}^\infty d r_\parallel 
(r_\parallel^2+r_\perp^2)_-^{\omega/2} e^{i q r_\parallel}
\label{-2<w<0.2}
\eea
where $(r_\parallel^2+r_\perp^2)_+^{\omega/2}$ denotes the value of
$(r_\parallel^2+r_\perp^2)^{\omega/2}$ on the right hand side of the
cut, and $(r_\parallel^2+r_\perp^2)_-^{\omega/2}$ its value on the
left hand side of the cut. Making the change of variable of
integration $r_\parallel=it$ on both sides of the cut, with $t$ real, we have
\beq
(r_\parallel^2+r_\perp^2)_\pm^{\omega/2}=e^{\pm i\omega\pi/2}(t^2-r_\perp^2)^{\omega/2} \,.
\label{-2<w<0.3}
\eeq
Thus we obtain
\bea
I(r_\perp) &=&-2\sin(\omega\pi/2)
\int_{r_\perp}^\infty d t
(t^2-r_\perp^2)^{\omega/2} e^{-q t}\,, \label{-2<w<0.4}\\
&=&
-2r_\perp^{\omega + 1}\sin(\omega\pi/2)
\int_{0}^\infty d\phi
(\sinh\phi)^{\omega + 1} e^{-q r_\perp\cosh\phi}\,,\nonumber\\
&&
\label{-2<w<0.5}
\eea
where in the second equality we changed variables of integration from
$t$ to $\phi$ via $t=r_\perp\cosh\phi$

Substituting this into our expression \rf{-2<w<0.1} for $U_0(\bq)$
gives
\bea
U_0(\qv) &=&2K a^{-\omega}\sin(\omega\pi/2)
\int_0^\infty d\phi \,(\sinh\phi)^{\omega + 1}
\nonumber\\
&&\times\int d^{d-1}r_\perp
r_\perp^{\omega + 1}e^{-q r_\perp\cosh\phi}\,, \label{-2<w<0.6}\\
&=&
\left(\frac{K}{a^{\omega}q^{2+\sigma}}\right)
2\sin(\omega\pi/2) S_{d-1} \Gamma(2+\sigma)
\nonumber\\
&&\times\int_0^\infty d\phi \, \frac{(\sinh\phi)^{\omega + 1}}{(\cosh\phi)^{2+\sigma}}
\label{-2<w<0.7}
\label{-2<w<0.8}
\eea
where we have reversed the order of integration over $\phi$ and
$\br_\perp$, performed the $d-1$ integral over $\br_\perp$, and used
the relation $\omega=2-d+\sigma$ in the second equality.

The remaining integral over $\phi$ in the resulting expression
converges for all $\omega>-2$ and $d>1$, and is tabulated\cite{GS}.
This is readily seen to again recover \rf{Udef} and \rf{UFT}, with the
result given by
\bea
C(\sigma,d) 
&=&\frac{2\pi^{(d-1)/2}\Gamma(2+\sigma)\Gamma(\oh(\omega+2))}
{a^\omega \Gamma(\oh(\sigma+3))}\;,\nonumber\\
&&
\label{c<sd}
\eea
where in the second equality we have again used \rf{sd-1}. 

Surprisingly, although this expression looks quite different from
\rf{csd}, it is, in fact, identical to it. To see this, we take the
ratio of the expression \rf{c<sd} for $C(\sigma,d)$ and the expression
\rf{csd} for it. Calling this ratio $R$, we find
\bea
R&=&\frac{2^\omega}{\omega \pi^{3/2}}\sin(\omega\pi)
\Gamma(\oh(\omega+2))\Gamma(1-\omega) \Gamma(\oh(\omega+1)) \;,
\nonumber\\
&&\\
&=&\frac{1}{\omega\pi}\sin(\omega\pi) 
\Gamma(1-\omega) \Gamma(\omega+1) \;,\\
&=&-\frac{1}{\pi}\sin(\omega\pi) 
\Gamma(-\omega) \Gamma(\omega+1) \;,
\label{R123}
\eea
where we used the general identity for Gamma functions\cite{GS} 
\beq
\Gamma(2x)={2^{2x-1}\over\sqrt{\pi}}\Gamma(x)\Gamma(x+1/2)
\label{Gamma}
\eeq
with $x=\oh(2+\sigma)$ and $x = \oh(1+\omega)$,
\bea
\Gamma(2+\sigma)&=&{2^{\sigma+1}\over\sqrt{\pi}}\Gamma\left({2+\sigma\over2}\right)\Gamma\left({3+\sigma\over2}\right)\,. \label{G1}\\
\Gamma(1+\omega)&=&\frac{2^\omega}{\pi^{1/2}}
\Gamma\left({1+\omega\over2}\right)\Gamma\left({2+\omega\over2}\right)\,,
\label{G2}
\eea
%
$\Gamma(1-\omega)=-\omega\Gamma(-\omega)$, and the identity
\beq
\Gamma(1-x)\Gamma(x)={\pi\over\sin(\pi x)}
\label{G3}
\eeq
with $x=1+\omega$, which reduces $R$ to 
\beq
R=-\left({\sin(\omega\pi)\over\pi}\right)\left({\pi\over\sin((\omega+1)\pi)}\right)=1 \,.
\label{Rf}
\eeq
This thereby proves that our expressions \rf{c<sd} and \rf{csd} are identical.

\subsubsection{$ \omega < -2$}

For $\omega<-2$, the approach of the previous subsection does not
work, since the integral over $\phi$ in Eq.\rf{-2<w<0.7} does not
converge.

We therefore take an alternative approach of calculating the integrals
for the Fourier transform from real space to Fourier space in
Eq.\rf{FTdir1} by working in (hyper-) spherical coordinates. After
taking $d-2$ azimuthal angular integrals in \rf{FTdir1}, we obtain
\beq
U_0(\bq) 
= -\frac{K}{a^\omega}S_{d-1}\int_0^\infty d r r^{\sigma+1}\int_0^\pi\sin^{d-2}\theta\; e^{-i q r\cos\theta}\;.
\label{w<-2.1}
\eeq
Since we are restricting ourselves here to $\sigma>-2$ and
$\omega<-2$, which implies $\sigma=d-2+\omega<d-4$, we see that
$-1<1+\sigma<d-3$. Hence, for $d\le4$, $-1<1+\sigma<0$. This is {\it
  precisely} the condition required to make the integral over $r$ in
\rf{w<-2.1} converge in {\it both} the infra-red ($r\to\infty$; here the
convergence is due in part to the oscillation of the complex
exponential) and the ultra-violet ($r\to0$). This constraint on the
exponent $1+\sigma$ also means we can close contours in the complex
plane as in Fig.\ref{contourIntegralFig}(a) for $\cos\theta>0$, and as
in Fig.\ref{contourIntegralFig}(b) for $\cos\theta<0$. This
calculation is so similar to the one we did subsection \rf{0<omega<1}
above that we will leave it to the reader to go through the details
here, and simply note that the result is again Eq.\rf{UFT},
with now:
\beq
C(\sigma,d)=\frac{S_{d-1}\cos(\oh\sigma\pi)\Gamma(2+\sigma)
  \Gamma(\oh(d-1))\Gamma(-\oh(1+\sigma))}{a^\omega\Gamma(-\oh\omega)}\;.
\label{w<-2.3}
\eeq
Once again, although this expression looks very different from
\rf{csd}, it is, in fact, identical to it. To see this, we take the
ratio of the expression \rf{w<-2.3} for $C(\sigma,d)$ and the
expression \rf{c<sd} for it (we proved earlier that \rf{c<sd} is
identical to \rf{csd}). Calling this ratio $R$, we find
\bea
R&=&\frac{\cos(\oh\sigma\pi)\Gamma(-\oh(1+\sigma))
  \Gamma(\oh(3+\sigma))}{\sin(\oh\omega\pi)\Gamma(\oh(1+\omega))
  \Gamma(\oh\omega)}\;.
\label{w<-2.4}
\eea
Using equation \rf{G3} in the numerator with $x=-\left({1+\sigma\over2}\right)$ shows that the numerator of this expression is equal to $-\pi$. Likewise, using equation \rf{G3} in the denominator with $x=-{\omega\over2}$ shows that the  denominator of this expression is also equal to $-\pi$. Hence, $R=1$, proving that \rf{w<-2.3} is identical to \rf{c<sd}, and, hence, to \rf{csd}.

\subsubsection{$d=1$, $-1<\omega<0$}

For a variety of reasons (e.g., the absence of components $\br_\perp$
of $\br$ perpendicular to $\bq$), the one-dimensional case has to be
treated separately for $\omega<0$.  Fortunately, for $-1<\omega<0$,
the necessary integrals to directly calculate $U_0(q)$ from $U_0(r)$
(note that both $r$ and $q$ are scalars in one dimension) can easily
be done by the sort of complex contour techniques we have used
above. We start with
\bea
U_0(q) &=& -\frac{K}{a^\omega}\int_{-\infty}^\infty d r
|r|^\omega e^{-i q r}\;,\\
&=& -\frac{K}{a^\omega}\int_0^\infty d r
r^\omega e^{i q r}+{\text{complex conjugate}}\;.\nonumber
\label{d=1.1}
\eea
For $-1<\omega<0$, the first integral on the second line converges at
both small and large $r$, and can be evaluated by rotating the contour
as in Fig.\ref{contourIntegralFig}(b). Making the change of variable
of integration from $r$ to $t$ via $r=it/q$ gives
\bea
U_0(q)&=& -\frac{K}{a^\omega q^{\omega+1}} 2\cos[(\omega+1)\pi/2]
\int_0^\infty d t t^\omega e^{-t}\;,\label{d=1.3}\\
&=& -\frac{K}{a^\omega q^{2+\sigma}} 2\cos[(\omega+1)\pi/2]
 \Gamma(\omega+1)\;. \label{d=1.4}
\eea
where in the second equality we have used $\omega=2-d+\sigma$ with
$d=1$.
This recovers Eq. (4) of the Introduction, with

\beq
C(\sigma,d=1) = 2\sin(\omega\pi/2)\Gamma(\omega+1)/a^\omega\,.
\label{d=1.5}
\eeq
Once again, although this constant looks quite different from its
expression in \rf{csd}, it is, in fact, identical to it, for $d=1$ and
$\omega=1+\sigma$.

\subsubsection{Potentials that fall off faster than $r^{-d}$}\label{dhapp}

Potentials that fall off faster than $r^{-d}$ are qualitatively
different from the cases just considered, since their volume integrals
converge at large $r$. Thus the volume integral of such potentials is
non-infinite, assuming that any short distance divergences are
regularized, either by introducing a lattice, or having a regularized,
integrable behavior of the potential at small $r$.

In contrast to the potentials with $\sigma>-2$ that we have been
considering up to now, such potentials therefore, have Fourier
transforms that remain {\it finite} as wavenumber $q\to0$.  The
``Debye-Huckel screened'' potentials treated in the main text have
this property, as we\Õll now show.

Here, we therefore consider the Fourier transform of a potential
$U(\br)$ which at large $r$ falls off like
\beq U(\br)\approx K_{DH}
\left({a\over r}\right)^\gamma
\label{ufast1}
\eeq
with $\gamma>d$, and is integrable at small $r$.

As noted above, although such a potential has a {\it finite} Fourier
transform $U(\bq)$ as $\bq\to0$, as we will now show, it always
exhibits non-analyticity at sufficiently high order in $q$. To see
this, we begin by noting that, due to the isotropy of the real space
potential $U(\br)$, the Fourier transform $U(\bq)$ is a function only
of the {\it magnitude} $q$ of $\bq$. Defining this function via
$U(\bq)\equiv U(q)$, and differentiating the Fourier transform
$U(\bq)$ of $U(\br)$ with respect to $q$ gives
\beq
\frac{dU}{dq} = - i K_{DH} a^\gamma S_{d-1}\int_0^\pi d\theta
\sin^{d-2}\theta\cos\theta I(q\cos\theta)\,,
\label{g1}
\eeq
where we have defined
\beq
I(q\cos\theta) = \int_0^\infty dr
r^{d-\gamma} e^{-i q r\cos\theta}
\label{g2}
\eeq
For $\gamma$ in the range $d+1>\gamma>d$, the integral in this
expression converges at non-zero $q$, but diverges as $q\to0$. We can
extract that divergence by rotating the contour as in
Fig. \rf{contourIntegralFig}(a) for $\cos\theta>0$, and as in
Fig. \rf{contourIntegralFig}(b) for $\cos\theta < 0$. Using those two
results in \rf{g1} gives
\begin{widetext}
\bea
\frac{dU}{dq} &=& - i \frac{K_{DH} a^\gamma S_{d-1}\Gamma(d-\gamma+1)}{q^{d-\gamma+1}}
\left[e^{-i\oh\pi(d-\gamma+1)}\int_0^{\pi/2} d\theta 
    \sin^{d-2}\theta\cos^{\gamma-d}\theta\right.\nonumber\\
  &+&\left.
e^{i\oh\pi(d-\gamma+1)}\int_{\pi/2}^{\pi} d\theta 
\sin^{d-2}\theta \cos\theta(-\cos\theta)^{\gamma-d-1}\right]\;,
\label{g3}\\
&=&
- K_{DH} a^\gamma q^{\gamma-d-1} 2\pi^{(d-1)/2}\cos(\pi(\gamma-d)/2)
\Gamma(d-\gamma+1) \Gamma(\oh(\gamma -d +1))/\Gamma(\gamma/2)\;,
\label{g4}
\eea
\end{widetext}
where to obtain the final equality, we substituted $\phi=\pi-\theta$ in
the second integral to reduce it to the minus the first integral,
which is tabulated. This result holds for $d+1>\gamma>d$.

To extend the Fourier transform result for values of $\gamma$ to the
range $d+n>\gamma>d+n-1$, where $n$ is an integer, we can proceed
similarly by taking $n$ derivatives of $U(q)$, which gives
\bew
\bea
\frac{d^nU}{dq^n} &=& (- i)^n K_{DH} a^\gamma S_{d-1}\int_0^\pi d\theta
\sin^{d-2}\theta\cos^n\theta\nonumber\\
&&\times \int_0^\infty dr\,  r^{d-\gamma -1 + n} e^{-i q
  r\cos\theta}\;, \label{g5}\\
&=&
(-1)^n K_{DH} a^\gamma q^{\gamma-d-n} 2\pi^{(d-1)/2}\cos(\pi(\gamma-d)/2)
\Gamma(d-\gamma+n) \Gamma(\oh(\gamma -d +1))/\Gamma(\gamma/2)\;,
\label{g6}
\eea
\ew
where the integral over $r$ converges by construction for all non-zero
$q$, and diverges as $q\to0$.  Integrating this expression $n$
times,  and making repeated use of the recursion relation for Gamma
functions $\Gamma(x+1)=x\Gamma(x)$, we obtain
\begin{widetext}
\bea
U(q) &=& \sum_{m=0}^{[n/2]}a_m q^{2m}
+\frac{  2\pi^{(d-1)/2}\cos(\pi(\gamma-d)/2)
\Gamma(d-\gamma) \Gamma(\oh(\gamma -d +1))}{\Gamma(\gamma/2)}
K_{DH} a^\gamma q^{\gamma-d}\;,\label{g7}
\eea
\end{widetext}
where the $a_m$\Õs are constants of integration characterizing the
analytic part of $U(q)$, $[x]$ is the greatest integer function, and
only even powers survive because all odd derivatives of $U(q)$ at
$q=0$ vanish, being proportional to integrals of odd powers of
$\cos\theta$.

In section \rf{dh}, we showed that the first non-analytic term in the
expansion of the Fourier tramsform $U_{\rm eff}(\bq)$ of the
Debye-Huckel screened potential $U_{\rm eff}(\br)$ is
$-\left({(\kbt)^2\tilde K\over g^2}\right)q^{2+\sigma}$. Comparing
this with our result above implies that we must have
$\gamma=d+2+\sigma$, which implies Eq.\rf{unconventionalDH}. Equating
the coefficients gives our expression \rf{cdhdef} for $C_{DH}$ with
\beq
G(\sigma,d)={\Gamma\left({d+\sigma+2\over2}\right)\pi^{1-d\over2}\over2\Gamma\left({3+\sigma\over2}\right)\cos\left({\sigma\pi\over2}\right)\Gamma(-2-\sigma)}
\,.
\label{Gsig}
\eeq
Plots of $G(\sigma,d)$ for $d=1,2$ and $3$ are given in Figs
\rf{gplot1}, \rf{gplot2}, and \rf{gplot3}, respectively.

\begin{figure}
   \centering
    \includegraphics[width=0.9\linewidth]{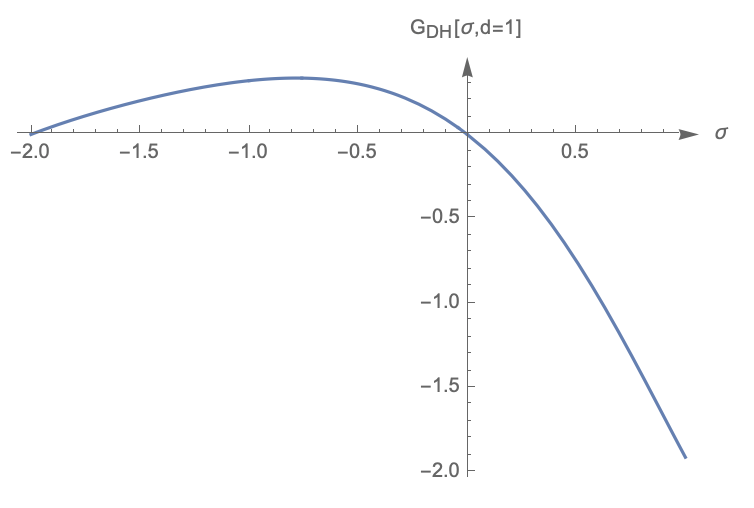}
    \caption{Plot of $G(\sigma,d=1)$ for the range of $\sigma$s
      accessible to our theory over which the potential can be
      Debye-Huckel screened, which is $-2<\sigma<1$. }
    \label{gplot1}
  \end{figure}
  
\begin{figure}
  \centering \includegraphics[width=0.9\linewidth]{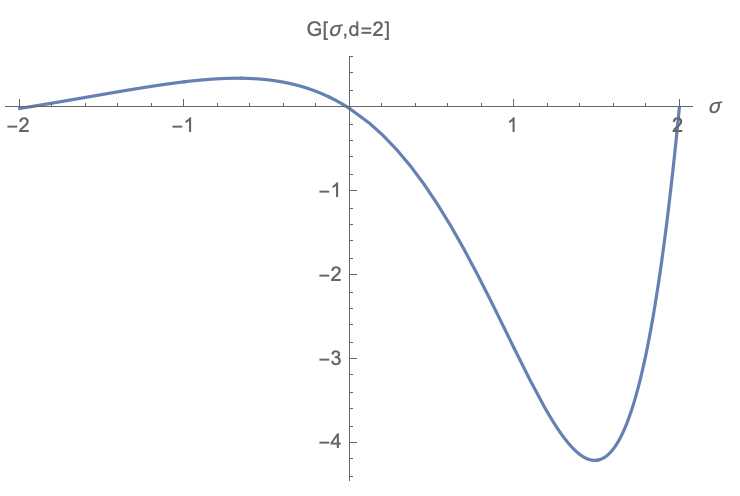}
  \caption{Plot of $G(\sigma,d=2)$ for the range of $\sigma$s
    accessible to our theory over which the potential can be
    Debye-Huckel screened, which is $-2<\sigma<2$. }
    \label{gplot2}
  \end{figure}
  
\begin{figure}
  \centering \includegraphics[width=0.9\linewidth]{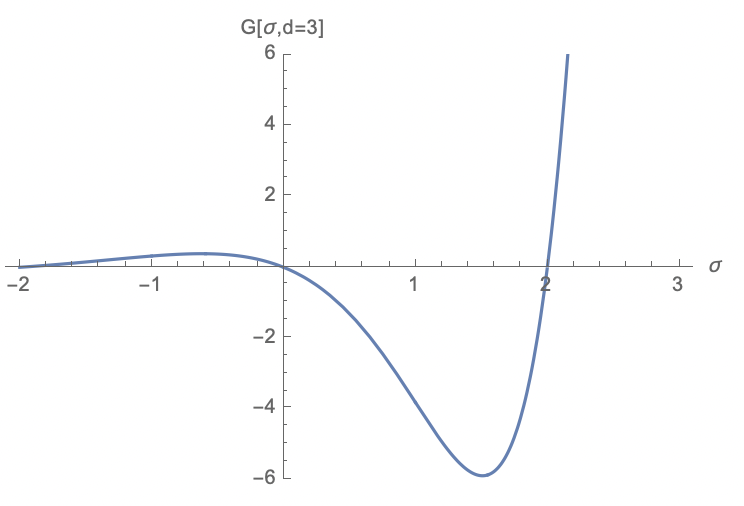}
  \caption{Plot of $G(\sigma,d=3)$ for the range of $\sigma$s
    accessible to our theory over which the potential can be
    Debye-Huckel screened, which is $-2<\sigma<3$.}
    \label{gplot3}
  \end{figure}
  The apparent singularities in \rf{Gsig} at $\sigma=\pm1$ are not, in
  fact, present, due to the cancellation of the divergences of
  $\Gamma(-2-\sigma)$ at those values of $\sigma$ with the vanishing
  of the $\cos\left({\sigma\pi\over2}\right)$ factor. Taking the limit
  $\sigma\to\pm1$ of \rf{Gsig} carefully, with a little help from the
  Marquis de l'H\^{o}pital, we find the perfectly finite results
\beq
G(\sigma=-1,d)={\Gamma\left({d+1\over2}\right)\over\pi^{d+1\over2}} \,,
\label{Gsig-1}
\eeq
and
\beq
G(\sigma=1,d)=-{6\Gamma\left({d+3\over2}\right)\over\pi^{d+1\over2}} \,.
\label{Gsig1}
\eeq
Indeed, $G(\sigma,d)$ is a perfectly smooth, non-singular function of
$\sigma$ for all three physical $d$\Õs, as can be seen from the plots
of figures \rf{gplot1}, \rf{gplot2}, and \rf{gplot3}.

\section{Summary of long-range Ginzburg-Landau model
  results}
\label{AppendixGL}
In this Appendix we summarize the results of a complementary
Ginzburg-Landau approach by Fisher, et al.\cite{Fisher}, which treats
the long-range interacting PM-FM transition from the paramagnetic
side. Working in terms of the soft-spin order parameter
$\psi = |\psi| e^{i\phi} = S_x + i S_y$,
\begin{equation}
  H_{\rm Fisher} =\frac{J}{2}\int_{\bf q}q^{2-\sigma}|\vec S_{\bf q}|^2 + u
  \int_{\bf r}(\vec S\cdot\vec S)^2\;,
\label{FisherGLlr}
\end{equation}
Fisher et.  al.\cite{Fisher} have analyzed the transition in a
long-ranged $O(n)$ model in $d$ dimensions with power-law
exchange\cite{translation}
\begin{equation}
  J_F(r)\sim \frac{1}{r^{d+\sigma_F}}\;,
    \label{fisherJ}
\end{equation}
where in terms of our exponent $\sigma$, their exponent
$\sigma_F = 2-\sigma$.  Applying their results to the 2D XY model, we
observe that the transition is clearly {\it not} of the
Kosterlitz-Thouless type that one would expect were the screening of
long-range exchange at play (it is {\em not}, as we argued in the main
text) via the (failed) mapping onto screened super-Coulomb gas
analyzed in the main text.

Indeed, Fisher et.al.'s work showed that in $d$-dimensions the
transition is characterized by the conventional critical exponents
$\eta$, $\nu$, $\gamma$, $\alpha$, and $\beta$, giving, respectively,
the behavior of the spin-spin correlation function at the transition:
\beq
\langle\bS_\rv\cdot\bS_{\rv'}\rangle\propto|\rv-\rv'|^{2-d-\eta} \,,
\label{expdef1}
\eeq
the divergence to the spin-spin correlation length $\xi$, the magnetic
susceptibility $\chi$, and the specific heat $C$ with temperature as
the transition temperature $T_c$ is approached:
\beq
\xi\propto|T-T_c|^{-\nu} \,\,\,,\,\,\,
\chi\propto|T-T_c|^{-\gamma}
\,\,\,,\,\,\,C\propto|T-T_c|^{-\alpha} \,,
\label{expdef2}
\eeq
and, last but not least, the vanishing of the magnetization
$M\equiv|\langle \bS_\rv\rangle|$ with $T_c-T$ as the transition is
approached from low temperature:
\beq M\propto|T_c-T|^{\beta} \,.
\label{betadef}
\eeq
Fisher et. al.\cite{Fisher,translation} found in $d=2$:
\beq
\eta=2-\sigma \,\,\,,\,\,\,\nu={1\over2-\sigma} \,\,\,,\,\,\,\beta={1\over2} \,\,\,,\,\,\,\gamma=1 \,\,\,,\,\,\,\alpha={2(1-\sigma)\over2-\sigma}
\label{expGauss}
\eeq
for $1<\sigma<2$ (the long-range analog of the mean-field behavior),
and
\begin{eqnarray}
\eta&=&1+{\epsilon\over 2}+O(\epsilon^2) \,\,\,,\,\,\,\nu=1+{2\epsilon\over5} +O(\epsilon^2)
  \nn\\
\beta&=&{1\over2}+{13\epsilon\over20}+O(\epsilon^2) \,\,\,,\,\,\,\gamma=1+{2\epsilon\over5} +O(\epsilon^2) \,\,\,,\,\,\,
  \nn\\
\alpha&=&-{4\epsilon\over5}+O(\epsilon^2)
\label{expnonG}
\end{eqnarray}
for $0<\sigma<1$, where $\epsilon\equiv2(1-\sigma)$.    

For the borderline case of $\sigma=1$ (which is relevant to the
experiments of Chen, et. al.,\cite{Browaeys2022}), one obtains universal
logarithmic corrections:
\bew
\beqn
\xi&\propto&|T-T_c|^{-1}\bigg[\ln\left(T_c\over{|T-T_c|}\right)\bigg]^{2/5}\propto\chi\nn\\
M&\propto&|T-T_c|^{1/2}\bigg[\ln\left(T_c\over{|T-T_c|}\right)\bigg]^{3/10}\,\,\,,\,\,\,C\propto\bigg[\ln\left(T_c\over{|T-T_c|}\right)\bigg]^{-4/5}\;.\nn\\
\label{logs}
\eeqn
\ew
None of these is remotely like the Kosterlitz-Thouless transition, for
which $\nu=\infty$ and $\alpha=-\infty$\cite{KT}. Hence, the
long-ranged XY model, despite the similarity of its single vortex
energy to that of our super-Coulombic plasma, does {\it not} belong to
the same universality class as the latter. In fact, it is clear that a
simple screening picture of vortices in the long-ranged XY model
fails.

Fortunately, such a picture is not needed, at least for $\sigma$
between $1$ and $2$, or $\sigma$ slightly less than $1$, for which the
above quoted results provide a complete picture. For $\sigma$ well
below $1$, there is no analytic approach we know of to obtain the
exponents. However, it seems exceedingly unlikely that the problem
will ever map onto the screened Coulomb problem we have considered in this
paper.


\begin{thebibliography}{99}
\bibitem{even} For even integers, this expansion fails because the
  Fourier transform of even power of $q$ is always a derivative of a
  delta function, and, hence, vanishes for non-zero $\br$. This
  vanishing is the result of the well-known exact cancellations that
  arise when one integrates the complex exponential $e^{i\bq\cdot\br}$
  over all $\br$. It does not occur once the complex exponential is
  multiplied by a non-even integer power of $q$.
\bibitem{range} The range of validity of this result is given in the
    Appendix\cite{SM}.
\bibitem{commentMap} A naive (i.e., incorrect) treatment of the
  long-range XY model in \rf{latticeH}, transforms it into a harmonic
  model \rf{modelH}, allowing vortices. Dualizing this model using
  vortex form in \rf{single vortex field} then seemingly transforms
  the power-law XY model to the power-law super/sub-Coulombic vortex
  gas, defined by $H$ in \rf{Hlr} and \rf{Udef}. Such transformation
  would thus erroneously predict a screening of the long-range XY
  model (with exchange in \rf{jdef} with $\sigma > 0$) to a
  conventional, short-range ($\sigma = 0$) XY model.
\bibitem{translation} Reference \cite{Fisher} uses a different
  definition of $\sigma$, which we refer to as $\sigma_F$, defining
  the exchange interaction to be
  $J_{\rv,\rv'}\propto {J_0\over|\rv-\rv'|^{d+\sigma_F}}$. Comparing this
  with our definition (generalized to arbitrary spatial dimensions
  $d$) $\propto { J_0\over|\rv-\rv'|^{d+2-\sigma}}$, we see that
  $\sigma_F=2-\sigma$, where $\sigma_F$ is the definition used in
  \cite{Fisher}, while $\sigma$ is the definition we use here. All of
  the discussion in our paper is in terms of our definition of
  $\sigma$, obtained by the substitution
  $\sigma\to \sigma_F \equiv 2-\sigma$ from the results of
  \cite{Fisher}.
\end{thebibliography}
\end{document}